%% file: EXO-16-033_temp.tex
\pdfoutput=1

\documentclass[11pt,twoside,a4paper,cmspaper,final,collab]{cms-tdr}
\def\svnVersion{467384P}\def\cmsCernNoTag{CERN-EP-2018-020}\def\cmsCernDate{\today}\def\cmsMessage{Published in the Journal of High Energy Physics as \href{http://dx.doi.org/10.1007/JHEP06(2018)128}{\doi{10.1007/JHEP06(2018)128.}}}

\begin{document}\cmsNoteHeader{EXO-16-033}

\hyphenation{had-ron-i-za-tion}
\hyphenation{cal-or-i-me-ter}
\hyphenation{de-vices}
\RCS$HeadURL: svn+ssh://svn.cern.ch/reps/tdr2/papers/EXO-16-033/trunk/EXO-16-033.tex $
\RCS$Id: EXO-16-033.tex 461251 2018-05-22 15:39:35Z hoepfner $

\newlength\cmsTabSkip\setlength{\cmsTabSkip}{1ex}

\providecommand{\MT}{\ensuremath{M_\mathrm{T}}\xspace}
\providecommand{\MTlower}{\ensuremath{M_\mathrm{T}^\mathrm{min}}\xspace}
\providecommand{\WprimeKK}{\ensuremath{\PW_{\mathrm{KK}}}\xspace}
\providecommand{\WprimeKKn}{\ensuremath{\PW^{\mathrm{(n)}}_{\mathrm{KK}}}\xspace}
\providecommand{\WprimeKKtwo}{\ensuremath{\PW^\mathrm{(2)}_{\mathrm{KK}}}\xspace}
\providecommand{\WMuNu}{\ensuremath{\PW \to \mu \nu }\xspace}
\providecommand{\tPZ}{\cPZ\xspace}
\providecommand{\tPW}{\PW\xspace}
\providecommand{\slep}[1][]{$\widetilde {\ell}$}

\cmsNoteHeader{EXO-16-033}
\title{Search for high-mass resonances in final states with a lepton and missing transverse momentum at $\sqrt{s}=13\TeV$}

\date{\today}

\abstract{
A search for new high-mass resonances in proton-proton collisions having final states with an electron or muon and missing transverse momentum is presented.
The analysis uses proton-proton collision data collected in 2016 with the CMS detector at the LHC at a
center-of-mass energy of 13\TeV, corresponding to an integrated luminosity of 35.9\fbinv.
The transverse mass distribution of the charged lepton-neutrino system is used as the discriminating variable.
No significant deviation from the standard model prediction is found.
The best limit, from the combination of electron and muon channels, is 5.2\TeV
at 95\% confidence level for the mass of a $\PWpr$ boson with the same couplings
as those of the standard model $\PW$ boson.
Exclusion limits of 2.9\TeV are set on the inverse radius of the extra dimension
in the framework of split universal extra dimensions.
In addition, model-independent limits are set on the production cross section and coupling strength of \PWpr bosons decaying into
this final state.
An interpretation is also made in the context of an R parity violating supersymmetric model
with a slepton as a mediator and flavor violating decay.
}

\hypersetup{%
pdfauthor={CMS Collaboration},%
pdftitle={Search for high-mass resonances in final states with a lepton and missing transverse momentum at sqrt(s) =13 TeV},%
pdfsubject={CMS},%
pdfkeywords={CMS, physics, W' decay}}

\maketitle

\section{Introduction}
\label{sec:intro}

The standard model (SM) of particle physics describes the properties of
all known elementary particles and the forces between them.
Five decades of experimental studies have verified its predictions to very high precision.
Despite the great success of the SM, theories beyond the SM (BSM) have been invoked to address a variety of open issues.
Many SM extensions predict additional heavy gauge bosons, including models with extended gauge sectors~\cite{PhysRevD.82.035011},
and theories with extra spatial dimensions~\cite{ArkaniHamed:1998rs, Appelquist:2000nn}.

The search presented in this paper is sensitive to deviations from the SM prediction in the
transverse mass spectrum of events with
a charged lepton (electron or muon) and a neutrino.
Interpretations of the observations are made in the context of several theoretical models:
the production and decay of a \PWpr boson in the sequential standard model (SSM)~\cite{reference-model},
the production and flavor violating decay of a slepton in an R-parity violating supersymmetry (RPV SUSY)
model~\cite{DreinerRPV, FarrarFayet},
and the production and decay of a Kaluza--Klein (KK) excitation of the \tPW boson in a model with split
universal extra dimensions (split-UED)~\cite{JHEP04(2010)081}.

The shape of the distribution is studied using a binned likelihood
method.
This approach is especially powerful as the examined theories predict different signal event distributions.
Although the details differ, all models predict
that the signal of a high-mass resonance is present at large transverse masses where SM backgrounds are very small.

The present analysis uses data corresponding to an integrated luminosity of 35.9\fbinv
of proton-proton collisions at a center-of-mass energy of 13\TeV, recorded in 2016 with the CMS detector
at the CERN LHC. The analysis improves upon the sensitivity of its predecessors~\cite{EXO-12-060, EXO-15-006},
benefiting from the increased energy and luminosity of the LHC.
Previous searches~\cite{ATLAS:2014wra, CONF-17-016} have not found evidence for
deviations from the SM prediction for the transverse mass distribution.

\section{The CMS detector}
\label{sec:detector}

The central feature of the CMS apparatus is a superconducting solenoid of 6\unit{m} internal diameter, providing a magnetic field of 3.8\unit{T}.
Within the solenoid volume are a silicon
pixel and strip tracker, a lead tungstate crystal electromagnetic calorimeter (ECAL), and a brass and scintillator hadron calorimeter (HCAL),
each composed of a barrel and two endcap sections.
Extensive forward calorimetry complements the coverage provided by the barrel and endcap detectors.

The silicon tracker measures charged particles within the pseudorapidity range $\abs{\eta}<2.5$. It consists of silicon pixel and silicon strip
detector modules.
The electromagnetic calorimeter consists of 75\,848 lead tungstate crystals that provide coverage in pseudorapidity $\abs{\eta}<1.48$
in a barrel region and $1.48<\abs{ \eta }<3.00$ in two endcap regions.
The ECAL energy resolution for electrons with a transverse momentum $\pt\approx45\GeV$ from $\Z \to \Pe \Pe$ decays is
better than 2\% in the central region of the ECAL barrel $(\abs{\eta}<0.8)$, and
is between 2\% and 5\% elsewhere~\cite{EGM-14-001}.
For high energies, which are relevant for this analysis, the electron energy resolution slightly improves~\cite{EXO-12-061}.

Muons are
measured in gas-ionization detectors embedded in the steel flux-return yoke outside the solenoid,
in the pseudorapidity range $\abs{\eta}<2.4$. Detection is provided using three technologies: drift tube (DT), cathode strip chamber (CSC),
and resistive plate chamber (RPC).
While the barrel region of $\abs{\eta} \leq1.1$ is instrumented with DT and RPC, the forward endcaps ($1.1<\abs{\eta}<2.4$)
are equipped with CSC and RPC.
A muon from the interaction point will cross four layers of muon chambers, interleaved with steel forming the return yoke of the magnetic field.
Every chamber provides reconstructed hits
on several detection planes, which are then combined into local track segments, forming the basis of muon reconstruction inside the muon system.
Matching muons track segments
to tracks measured in the silicon tracker results in a relative transverse momentum (\pt) resolution in the barrel of
about 1--2\% for muons with $\pt \lesssim200\GeV$ and
better than 10\% for high momentum muons of $\pt \sim1\TeV$~\cite{MuonPerformance13TeV}.

Jets are reconstructed offline from the energy deposits in the calorimeter towers, clustered using the anti-\kt
algorithm~\cite{Cacciari:2008gp, Cacciari:2011ma} with a distance parameter of 0.4. In this process, the contribution from each
calorimeter tower is assigned a momentum, the absolute value and the direction of which are given by the energy measured in the tower,
and the coordinates of the tower. The raw jet energy is obtained from the sum of the tower energies, and the raw jet momentum by the
vectorial sum of the tower momenta, which results in a nonzero jet mass. The raw jet energies are then corrected to establish a
relative uniform response of the calorimeter in $\eta$ and a calibrated absolute response in transverse momentum \pt.

The CMS experiment has a two-level trigger system~\cite{Khachatryan:2016bia}. The level-1 trigger, composed of
custom hardware processors, selects events of interest using information from the calorimeters
and muon detectors and reduces the readout rate from the 40\unit{MHz} bunch-crossing frequency
to a maximum of 100\unit{kHz}. The software based high-level trigger uses the full event
information, including that from the inner tracker, to reduce the event rate to the 1\unit{kHz} that is
recorded.

A more detailed description of the CMS detector can be found in Ref.~\cite{Chatrchyan:2008zzk}.

\section{Physics models and event simulation}
\label{sec:models}

Many BSM scenarios predict the existence of new particles that decay with the experimental signature of a charged
lepton, $\ell$, and missing transverse momentum, \ptmiss, where the latter may flag the presence of a non-interacting particle.
The missing transverse momentum, \ptvecmiss, is defined as ${-}\sum \ptvec$ of all reconstructed particles, with \ptmiss being the modulus of \ptvecmiss.

The analysis is performed in two channels: $\Pe+\ptmiss$ and $\mu+\ptmiss$.
New particles may be detected as an excess of events in the observed spectrum of transverse mass, defined as
\begin{equation}
\MT = \sqrt{2  \pt^\ell ~~ \ptmiss  \bigl(1 - \cos [\Delta \phi(\ell,\ptvecmiss)]\bigr) },
\label{eqn:mt}
\end{equation}
where $\Delta \phi(\ell,\ptvecmiss)$ is the azimuthal opening angle (in radians) between the directions of the missing transverse momentum and
the charged lepton.
Several new-physics models predict the production of high-\pt leptons, which should be identifiable as an excess
in the region of high \MT values where little SM background is expected.

This section summarizes the new physics models used for interpretation of the observations, along with model-specific assumptions and
details of the generator programs used
for production of simulated signal event samples. All generated events are processed through a full simulation of the CMS detector
based on
\GEANTfour~\cite{Agostinelli:2002hh}, a trigger emulation, and the event reconstruction chain.

All simulated event samples are normalized to the integrated luminosity of the recorded data.
The simulation
of pileup is included in all event samples by superimposing simulated minimum bias interactions onto all simulated events.
For the data set used, the average number of interactions per bunch-crossing after selection
is about 20, with a maximum of 55.

\begin{figure}[hbtp]
\centering
\includegraphics[width=.35\textwidth]{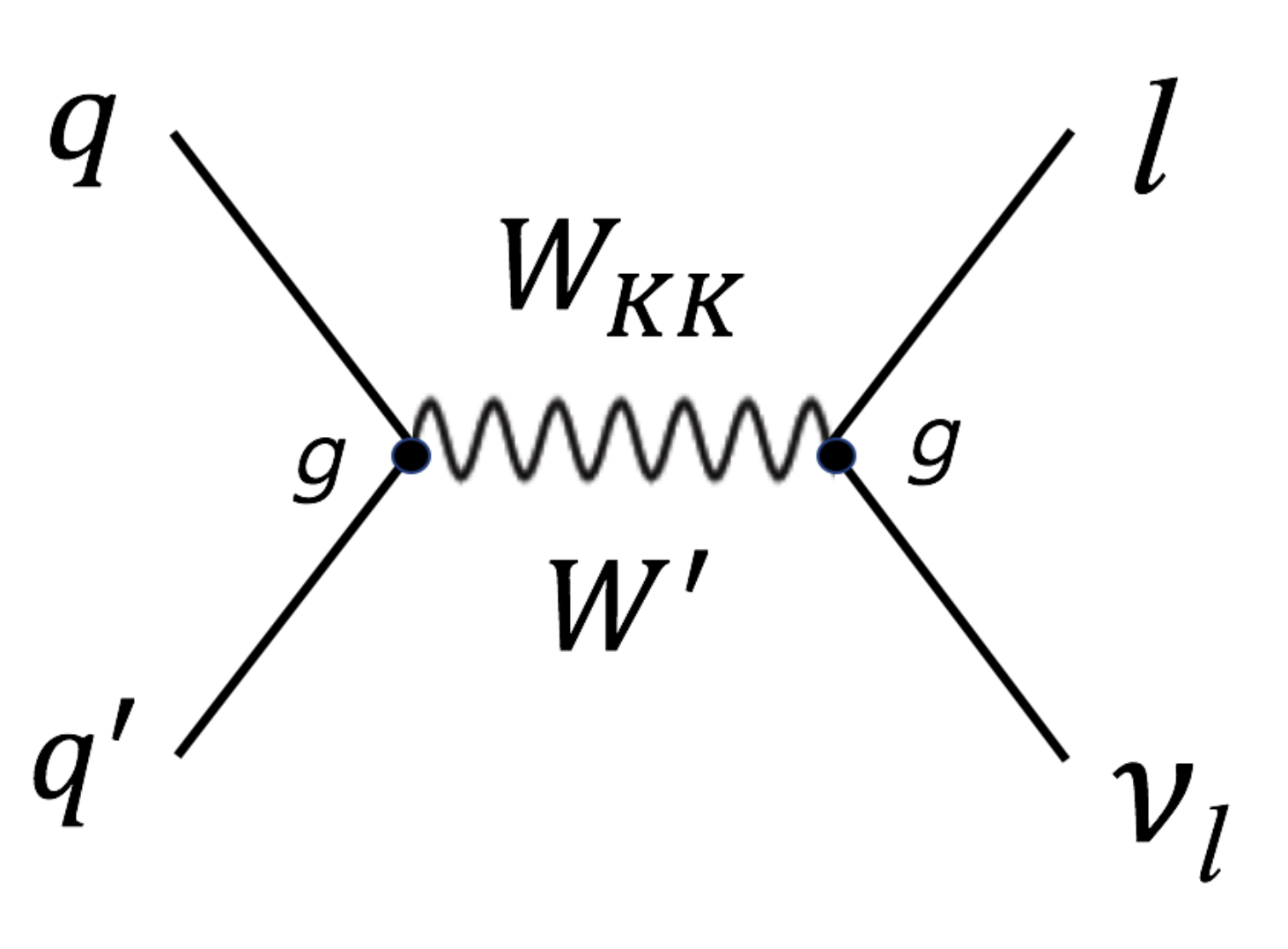}
\includegraphics[width=.4\textwidth]{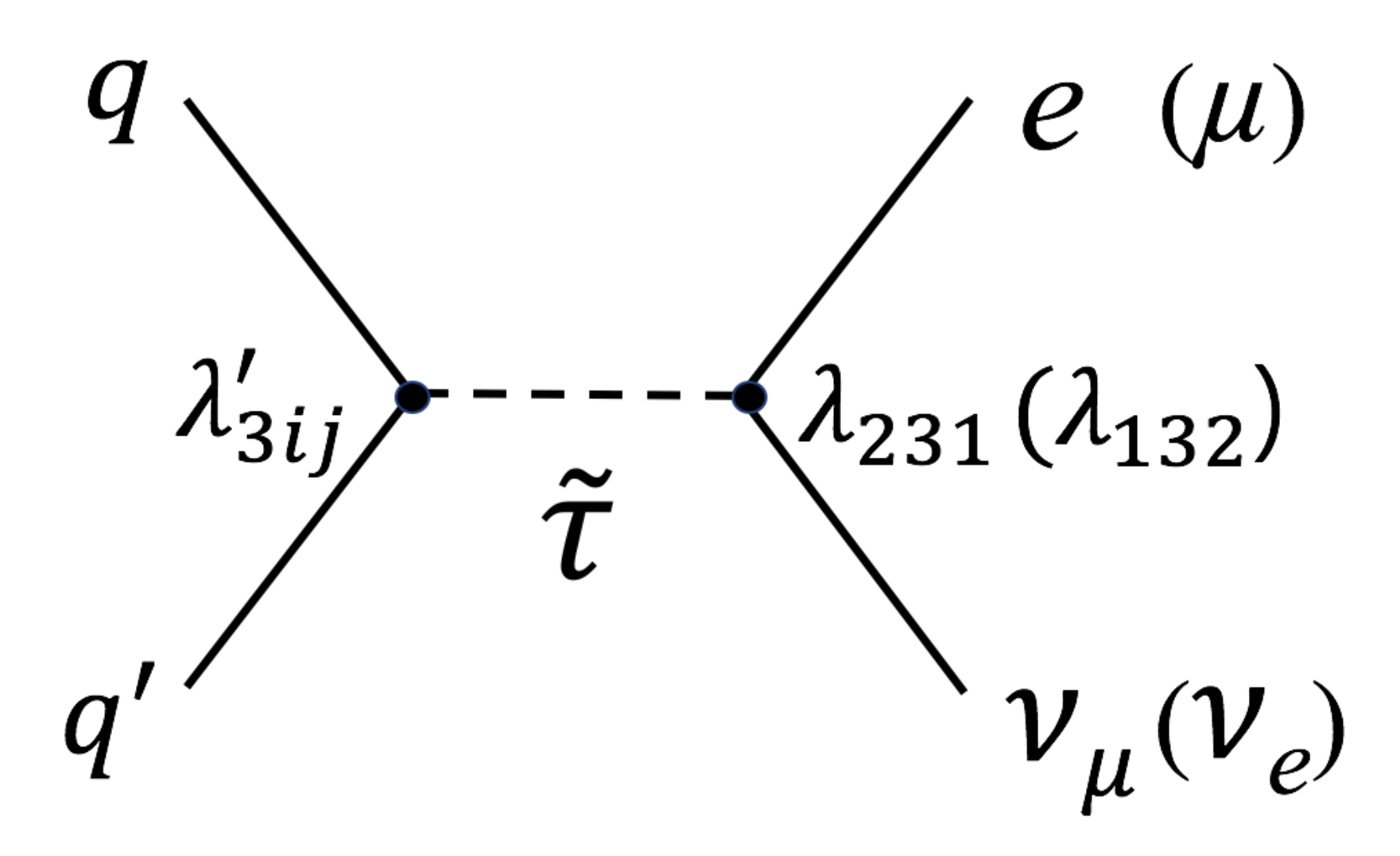}
\caption{Production and decay of a new heavy boson, an SSM \PWpr or a \WprimeKK (left). The coupling strength, $g$, may vary.
In RPV SUSY, a tau slepton (\sTau) could
also act as a mediator (right) with corresponding $\lambda$ couplings for the decay that are different for the two final states,
denoted by $\lambda_{231}$
and $\lambda_{132}$ for the electron and muon final states, respectively.
}
\label{fig:feynman}
\end{figure}

\subsection{Sequential standard model \texorpdfstring{\PWpr}{W'} boson}
\label{sec:models-ssm}

The SSM~\cite{reference-model} has been used as a benchmark model for experimental \PWpr boson searches
for more than two decades. The Feynman diagram for the production and decay of a \PWpr boson
is depicted in Fig.~\ref{fig:feynman} (left).
In accordance with previous analyses, no interference with the SM \tPW boson is considered.

In the SSM, the \PWpr boson is considered to be a heavy analogue of the SM \tPW boson, with similar decay modes and branching fractions.
These are modified by the presence of the
$\cPqt\cPaqb$ decay channel, which opens up for \PWpr boson masses above 180\GeV.
Dedicated searches in the $\cPqt\cPaqb$ channel are described in
Refs.~\cite{tbanalysis-d0, WprimeTBCMS, Aad:2014xra}.
For this search, the given assumptions yield a predicted branching fraction ($\mathcal{B}$) of
about 8.5\% for each of the leptonic channels studied. The width of a 1\TeV \PWpr boson would be about 33\GeV.
Decays of the \PWpr boson via \PW\tPZ are assumed to be
suppressed. Dedicated searches for these decays can be found in Refs.~\cite{Khachatryan:2014xja, CMS-WZ-2017}.

The signature of a \PWpr boson is a Jacobian peak in the transverse mass distribution,
similar to that of the SM \tPW boson, but at a higher mass.
Because of constraints from the parton distribution functions (PDF) the phase space for production of very massive \PWpr bosons in $\Pp\Pp$ collisions at 13\TeV is reduced,
leading to a large fraction of such \PWpr bosons being produced off-shell, at lower masses.

The simulation of data samples in the SSM is performed at leading order (LO) with \PYTHIA 8.212~\cite{Sjostrand:2014zea},
using the NNPDF2.3 PDF set~\cite{NNPDF2.3, NNPDF23} and tune CUETP8M1~\cite{CUETP8M1}.
The simulated masses range from 400\GeV to 6\TeV, where the lower mass matches
the beginning of the sensitive region as determined by the trigger thresholds.
A \PWpr boson mass-dependent K-factor is used to correct for next-to-next-to-leading order (NNLO)
QCD multijet cross sections, calculated using \FEWZ~3.1~\cite{fewz, Li:2012wna}.
The K-factor varies from 1.363 to 1.140.

\subsection{Varying coupling strength}

The \PWpr boson coupling strength, $g_{\PWpr}$,
is usually given in terms of the SM weak coupling strength $g_{\tPW}$.
If the \PWpr is a copy of the SM \tPW boson, their coupling ratio is $g_{\PWpr}/g_{\tPW} = 1$
and the SSM \PWpr theoretical cross sections, signal shapes, and widths apply.
However, different couplings are possible.
Because of the dependence of the width of a particle on its coupling, and the consequent effect on the \MT distribution,
a limit can also be set on the coupling strength.
For this study, signal samples for a range of coupling ratios, $g_{\PWpr}/g_{\tPW} = 10^{-2}$ to $3$, are simulated in LO with
\MADGRAPH~5 (v1.5.11)~\cite{madgraph5}.
These signals exhibit different widths as well as different cross sections.
The generated distributions of the \PYTHIA samples are reweighted to take into account the decay width dependence,
thus providing the appropriate reconstructed \MT distributions.
For $g_{\PWpr}/g_{\tPW} = 1$ the theoretical LO cross sections apply
and this coupling strength was used to compare the standard SSM samples with the reweighted ones,
allowing the reweighting method to be verified.
For $g_{\PWpr}/g_{\tPW}\neq1$ the theoretical cross sections
scale with the coupling strength squared.

\subsection{Split-UED model}
\label{sec:models-ued}

The leptonic final states under study may also be interpreted in the framework of universal
extra dimensions with fermions propagating in the bulk, known as split-UED model~\cite{PhysRevD.79.091702, JHEP04(2010)081}.
This is a
model based on an extended space-time with an additional compact fourth spatial dimension
of radius $R$, and a bulk mass parameter of the fermion field in five dimensions, $\mu$.
In this model all SM particles have corresponding Kaluza-Klein (KK) partners,
for instance \WprimeKKn,
where the superscript denotes the n$^{\mathrm {th}}$ KK excitation mode. Only KK-even KK
modes of \WprimeKKn couple to SM fermions, owing to KK-parity conservation.

In the split-UED model the parameter $\mu$ is assumed to be non-zero,
following Refs.~\cite{PhysRevD.79.091702, JHEP04(2010)081}.
The mass of the \WprimeKKn
is determined by M(\WprimeKKn) = $\sqrt{\smash[b]{M_{\PW} + (n/R)}}$,
i.e., a larger radius ($R$) corresponds to smaller KK masses. The mass of KK fermions
depends on the bulk mass parameter $\mu$. The product of the cross section of the \WprimeKKn
production and the
branching fraction to standard model fermions goes to zero as $\mu$ goes to zero.

For the mode $n=2$, the decay of \WprimeKKtwo
to leptons is kinematically identical to the SSM \PWpr boson
decay, and the observed limits obtained from the \PWpr$\to \Pe\nu$ and \PWpr$\to\mu\nu$ searches
can be reinterpreted directly in terms of the \WprimeKKtwo boson mass, taking into account the difference
in widths in the simulation.
The Feynman diagram in Fig.~\ref{fig:feynman} (left) shows this process.
The simulation is performed
at leading order with \PYTHIA 8.212. The mass-dependent K-factors from the SSM \PWpr interpretation are used.
This is possible since the signal shapes of a \WprimeKKtwo and a SSM \PWpr correspond to each other. The signal samples are generated
with the parameters $1/R = 200$ to 3000\GeV and $\mu = 50$ to 10\,000\GeV. The $1/R$ range corresponds to the mass of
\WprimeKKtwo from approximately 400 to 6000\GeV.

\subsection{RPV SUSY with scalar lepton mediator}
\label{sec:models-rpv}

This model assumes a SUSY scalar lepton (\slep) as a mediator, with subsequent R-parity and
lepton flavor violating decay to a charged lepton and a neutrino~\cite{DreinerRPV, FarrarFayet}.
The analysis studies the cases where a tau slepton decays to $\Pe + \; \nu_{\mu}$ or to $\mu + \nu_{\Pe}$.
The Feynman diagram is depicted on the right of Fig.~\ref{fig:feynman} for the two decay channels under study.
While on the production side, the coupling is always a version
of the hadronic-leptonic RPV coupling $\lambda^{\prime}_{3ij}$
(which is the coupling to the third generation, in this case the tau slepton),
the decay is governed by the leptonic RPV coupling $\lambda_{231}$ for the decay to $\Pe+\nu_{\mu}$,
and by $\lambda_{132}$ for the decay to $\mu+\nu_e$.
The values of the couplings may be identical.
Signal samples for a range of tau slepton masses, M(\sTau),
are simulated with \MADGRAPH~5 (v1.5.14) at LO and no higher order effects are considered.
Signals are simulated with the parameters $\lambda_{231}$, $\lambda_{132}$ and
$\lambda^{\prime}_{3ij}$=0.05 to 0.5 for
$M(\sTau)=400$ to 6000\GeV.

\section{Event reconstruction}
\label{sec:objects}

The models described in the previous section provide an event signature of a single high-\pt
lepton (electron or muon) and a particle that cannot be detected directly, giving rise to the
experimentally observed \ptmiss.
This quantity is measured using a particle-flow (PF) technique~\cite{PFT-2017paper},
that combines measurements from all components of the CMS detector in order to
produce particle candidates. The modulus of the vector \pt sum of these candidates defines \ptmiss,
which is corrected for the jet energy
calibration~\cite{PAS-JME-16-004,Chatrchyan:2011tn}. At high mass, the \ptmiss is mainly determined by the high-\pt lepton in the event.

Electrons are reconstructed as ECAL clusters that are matched to a central track and their
identification has been optimized for high-\pt values~\cite{EXO-15-005}.
Electron candidates are required to be isolated, have an electron-like shape, and be within the acceptance region of the barrel ($\abs{\eta}<1.44$)
or the endcaps ($1.56<\abs{\eta}<2.50$).
This acceptance region avoids the transition region between barrel and endcap parts of the ECAL.
Electron isolation in the tracker is ensured
by requiring the sum of \pt to be less than 5\GeV for all tracks that are in close proximity
to the track of the electron candidate and to originate from the same primary vertex.
Only tracks that are within a cone of $\Delta R=\sqrt{\smash[b]{(\Delta \phi)^2 + (\Delta \eta)^2}}<0.3$ around the electron candidate's track are considered.
The reconstructed vertex with the largest value of summed physics-object $\pt^2$ is taken to be the primary $\Pp\Pp$ interaction vertex. The physics objects are the jets, clustered using
the jet finding algorithm~\cite{Cacciari:2008gp,Cacciari:2011ma} with the tracks assigned to the vertex as inputs, and the associated missing transverse momentum, taken as the negative
vector sum of the \pt of those jets.
As in the tracker isolation calculation, in the calorimeters the sum of energy deposits within a cone $\Delta R<0.3$ around the electron candidate's direction is used as a measure of isolation.
It is corrected for the mean energy contribution from additional $\Pp\Pp$ collisions occurring within the same bunch crossing (pileup)~\cite{EGM-13-001}.
To obtain sufficiently isolated electrons, this calorimeter isolation is required to be below a threshold of 3\% of the electron's transverse momentum.
Additionally, the energy deposits in the hadron calorimeter within a cone of $\Delta R=0.15$ around the electron's direction must be less than 5\% of the electron's energy deposit in the ECAL.
In order to differentiate between electrons and photons, properties of the track matched to the calorimeter measurement must be consistent with those of an electron
originating from the primary vertex.
Specifically, there must be $\leq$1 hit missing in the innermost tracker layers, and the transverse distance to the primary vertex must be $<$0.02\unit{cm} (barrel) or $<$0.05\unit{cm} (endcap).

The muon system covers the pseudorapidity region $\abs{\eta}<2.4$.
The reconstruction of muons is optimized for high-\pt values~\cite{EXO-15-005}.
Information from the inner tracker and the outer muon system are used together.
Each muon track is required to have at least one hit in the pixel
detector, at least six tracker layer hits, and segments with hits in two or more muon detector stations. Since segments are typically
in consecutive layers separated by thick layers of steel,
the latter requirement significantly reduces the amount of hadronic punch-through~\cite{Chatrchyan:2012xi}.
To reduce background from cosmic ray muons, each muon is required to have a transverse impact parameter of less than 0.02\unit{cm}
and a longitudinal distance parameter of less than 0.5\unit{cm}
with respect to the primary vertex. In order to suppress muons with mismeasured \pt, an additional requirement $\sigma_{\pt}/\pt<0.3$ is applied, where
$\sigma_{\pt}$ is the \pt uncertainty from the muon track reconstruction.
Muon isolation requires that the scalar \pt sum of all tracks originating from the interaction
vertex within a $\Delta R<0.3$ cone around its direction, excluding the muon itself, is less than 10\% of the muon's \pt.

In order to determine any differences in the selection efficiencies between observed and simulated data, the
efficiencies for both channels are measured using the "tag-and-probe" method~\cite{CMS:2011aa}, with samples
of dilepton events from high-\pt \PZ boson decays and high-mass Drell-Yan pairs. The overall
efficiency in each case includes contributions from the trigger and the lepton reconstruction and identification
criteria. The ratio of the efficiencies for data and simulation, denoted as the scale factor (SF), is determined
for each channel separately. The SFs that match the electron identification and reconstruction
efficiencies in data and simulation are $0.972 \pm 0.006$ (barrel) and $0.983 \pm 0.007$ (endcap).
The muon scale factors are also sensitive to differences between simulated samples and data,
caused by radiative processes associated with muon interactions in the material
of the detector.
The corresponding SFs are applied as a function of $\eta$,
with uncertainties covering possible lower efficiencies in data and
 dependent on the muon momentum that, for a muon momentum of 5\TeV, range from 3\% for $\abs{\eta}<1.6$ and up to 20\% for $1.6<\abs{\eta}<2.4$.

\section{Event selection}
\label{sec:selection}

The event selection follows the approach used in previous CMS analyses~\cite{EXO-12-060, EXO-15-006}.
Events are triggered with a single-electron (muon) trigger with trigger thresholds of lepton $\pt>115(50)\GeV$.
In addition, high-\pt electrons may be triggered by a single-photon trigger
with a threshold of 175\GeV for the photon \pt,
which is used in a logical "OR" with the single-electron trigger.
For the electron channel, the trigger SF is determined to be
$0.989 \pm 0.003$ (barrel) and $0.996 \pm 0.003$ (endcap), respectively, for the combination of the single-electron and single-photon triggers.
The muon trigger SF accounts for observed differences between the efficiencies in simulation and in data for particular regions of the detector, such as the transition region
between the barrel and endcap muon
detectors. The SF values, applied as a function of $\eta$, range from 0.92 to unity.

\begin{figure}[hbtp]
\centering
 \includegraphics[width=0.49\textwidth]{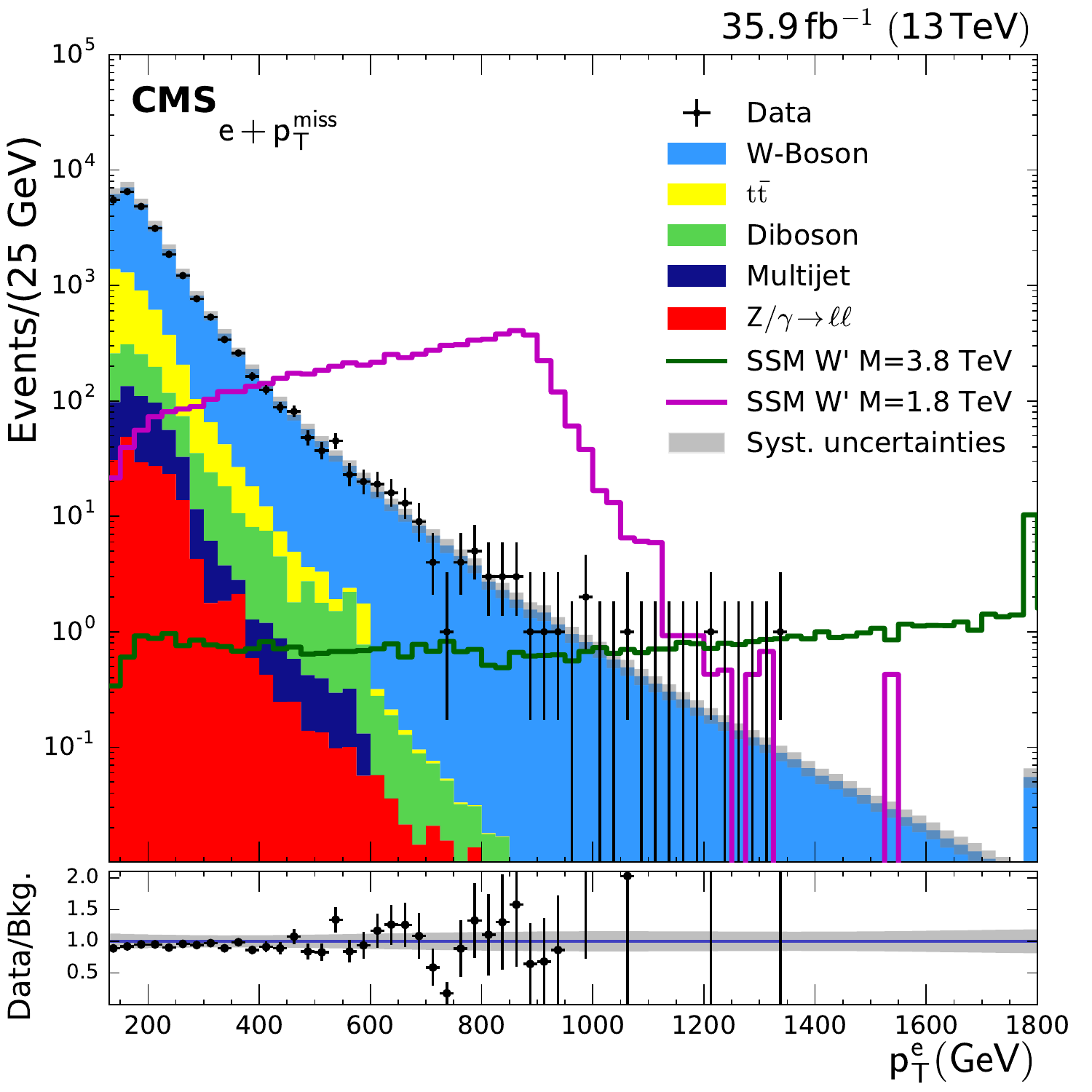}
 \includegraphics[width=0.49\textwidth]{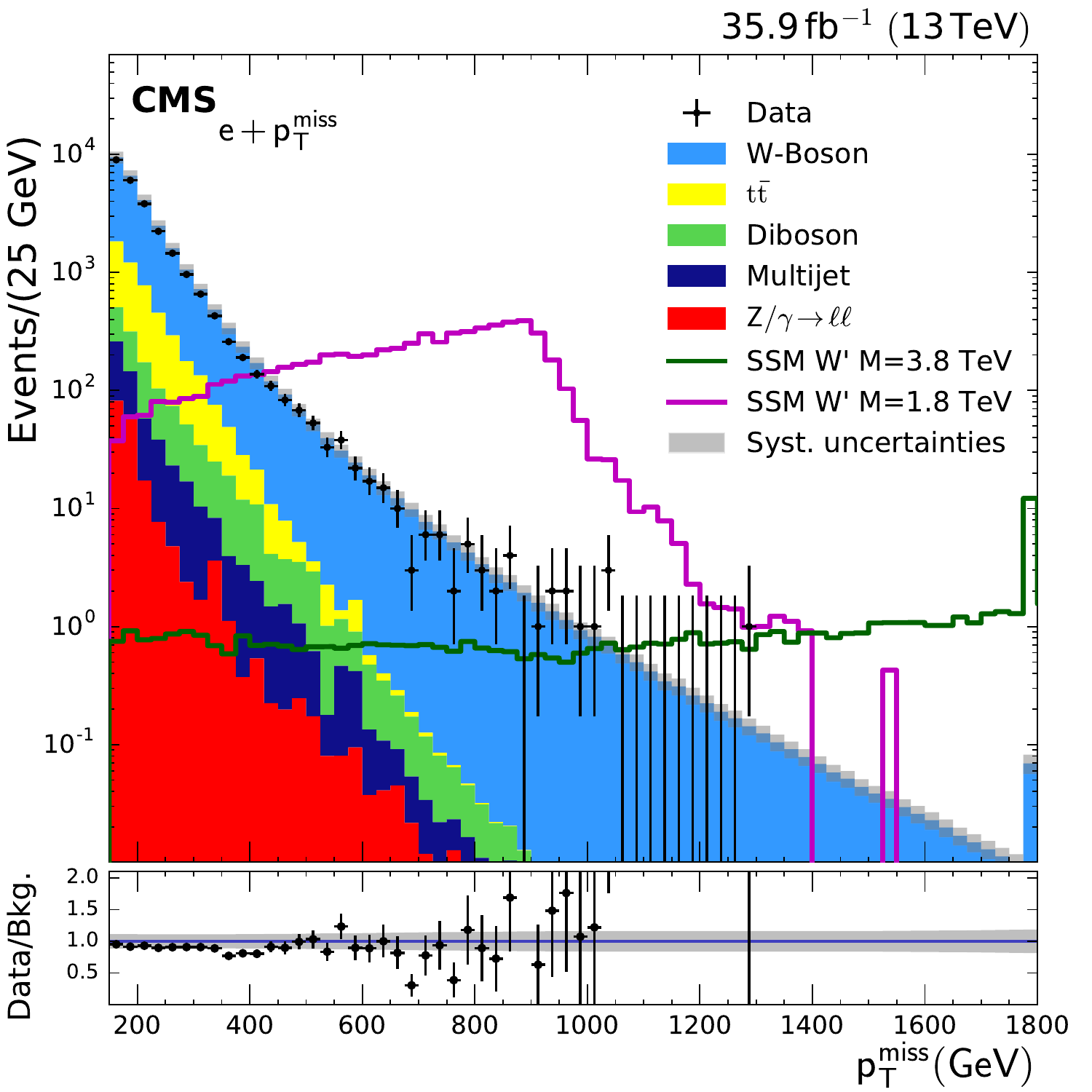}
 \includegraphics[width=0.49\textwidth]{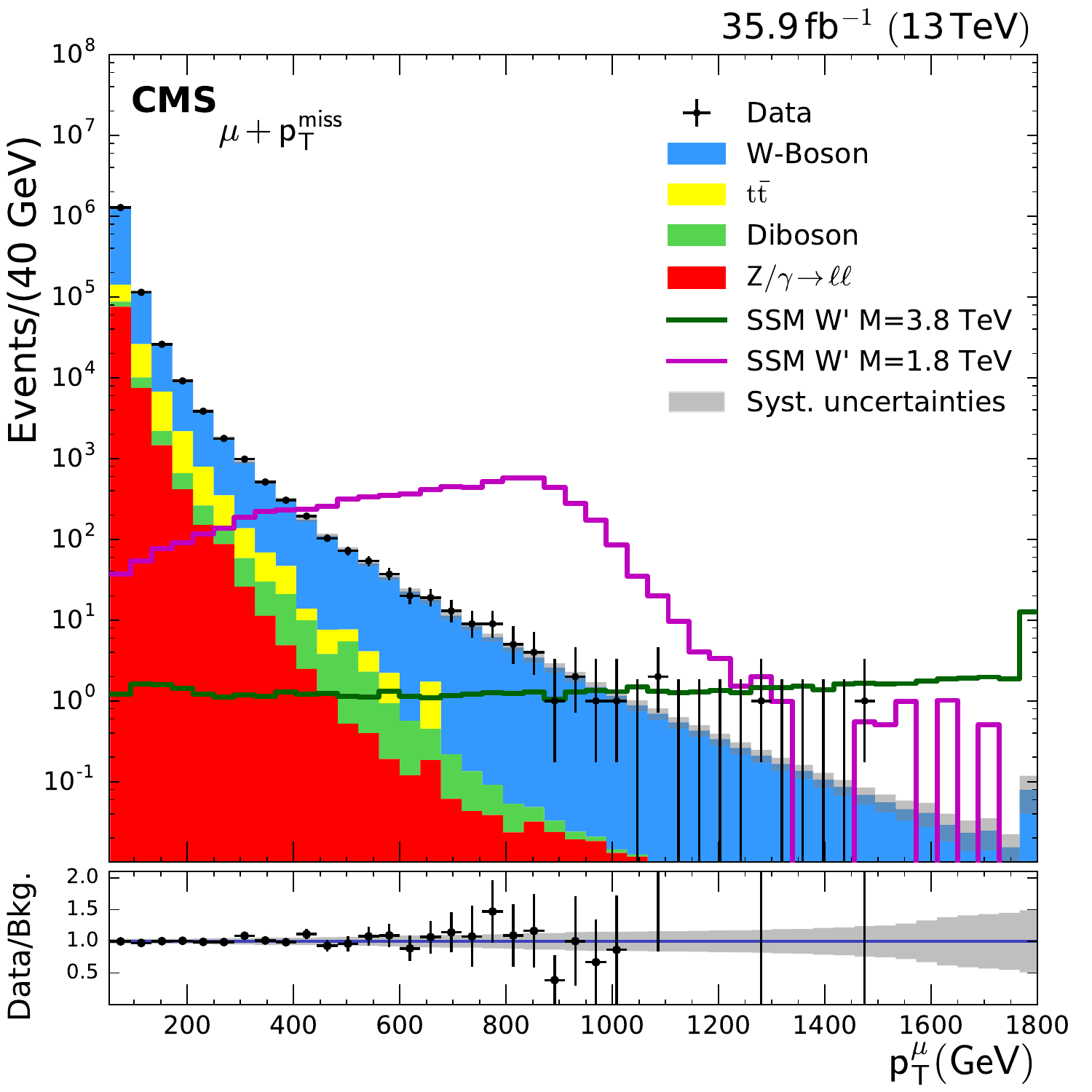}
 \includegraphics[width=0.49\textwidth]{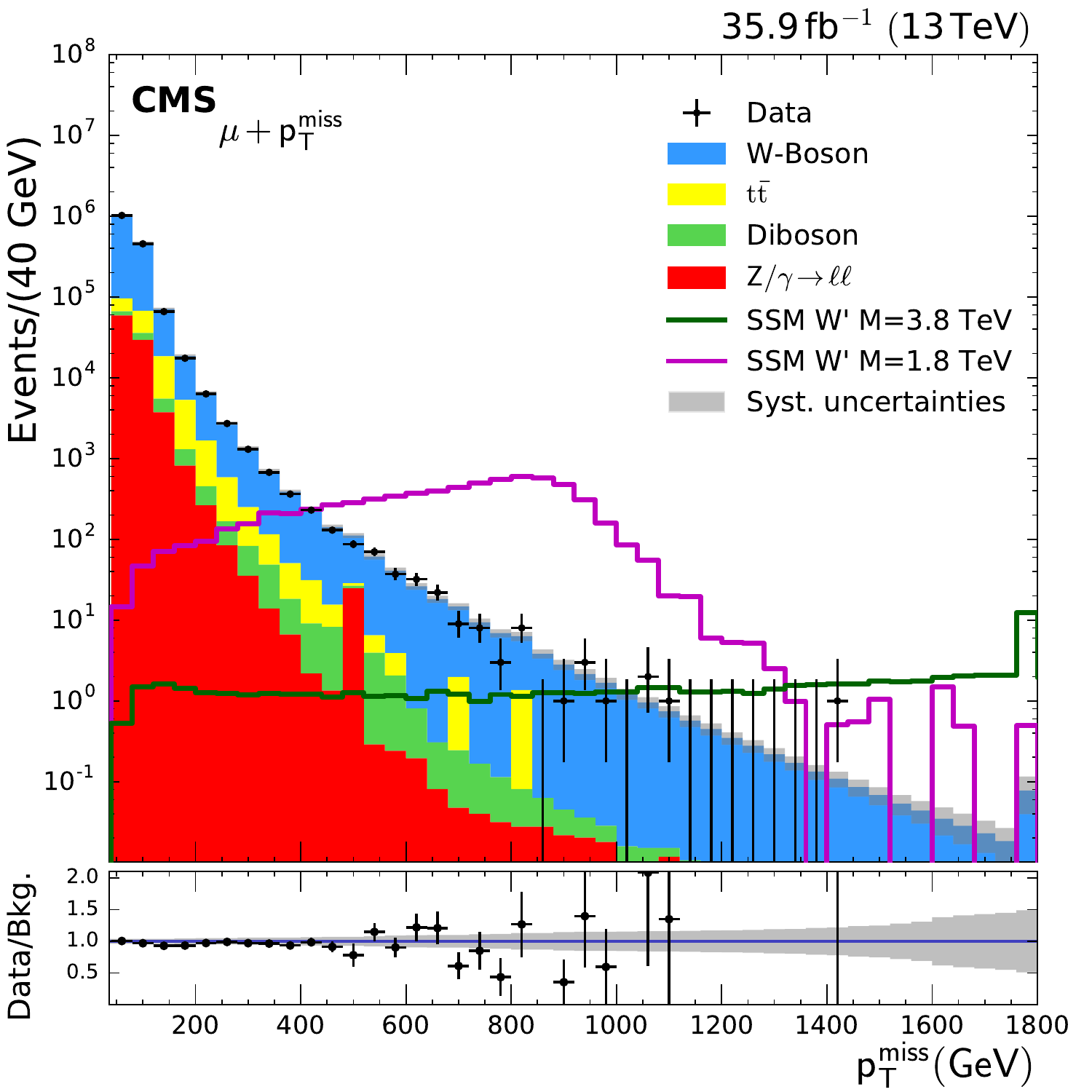}
\caption{Distributions in \pt (left) and \ptmiss (right), for
the electron (upper row) and muon (lower row)
for data and for expected SM backgrounds, after applying complete selection criteria.
The QCD multijet background in the electron channel is derived from data.
The background labelled as "diboson" includes $\PW\PW$, $\PZ\PZ$, and $\PW\tPZ$ contributions.
Also shown are SSM \PWpr signal examples for the two indicated masses.
The last bin shows the total overflow.
The lower panel shows the ratio of data to prediction and the shaded band
includes the systematic uncertainties.
}
\label{fig:BasicDistributions}
\end{figure}

The minimum offline lepton \pt must be above the lepton trigger threshold.
For offline reconstruction, electrons are required to have $\pt>130\GeV$ and to have $\abs{\eta}<1.44$
or $1.56<\abs{\eta}<2.50$. Muons must have $\pt>53\GeV$ and $\abs{\eta}<2.4$. Signal events are identified by the presence of an isolated high-\pt lepton. Events with a second
lepton with $\pt>25\GeV$ are rejected. In the electron channel, \ptmiss is required to be above 150\GeV to avoid the mismodeled low-\ptmiss region.

In the muon channel, the contribution from \ttbar events is further reduced by
rejecting events containing six or more jets or events with the leading jet having $\pt>25\GeV$ and $\abs{\eta}<2.5$, consistent with originating from a bottom quark, using the
standard CMS b-tagging tools~\cite{BTV-16-002}.

In the considered models, the lepton and \ptvecmiss are expected to be nearly back-to-back in the transverse
plane, and balanced in transverse momentum.
To incorporate these characteristics in the analysis, additional kinematic criteria select events based on
the ratio of the lepton \pt to \ptmiss, requiring $0.4<\pt/{\ptmiss}<1.5$, and on the angular
difference between the lepton  and \ptmiss, with $\Delta \phi(\ell,\ptvecmiss)>2.5 \approx0.8\pi$.
The distributions of the lepton \pt and \ptmiss are depicted in Fig.~\ref{fig:BasicDistributions}
separately for the electron (upper panels) and muon (lower panels) channels.

For simulated events passing all the selection criteria, the signal efficiency
for an SSM \PWpr with no requirement on the reconstructed \MT in the event,
is maximal at a value of 0.75 (for both decay channels) for a \PWpr boson mass range 1.5--2.5\TeV, and decreases gradually to $\approx$0.60
for larger and smaller masses.
For larger masses, the increasing off-shell production displaces events to lower \MT,
where the background is larger.

\begin{figure}[hbtp]
\centering
 \includegraphics[width=0.49\textwidth]{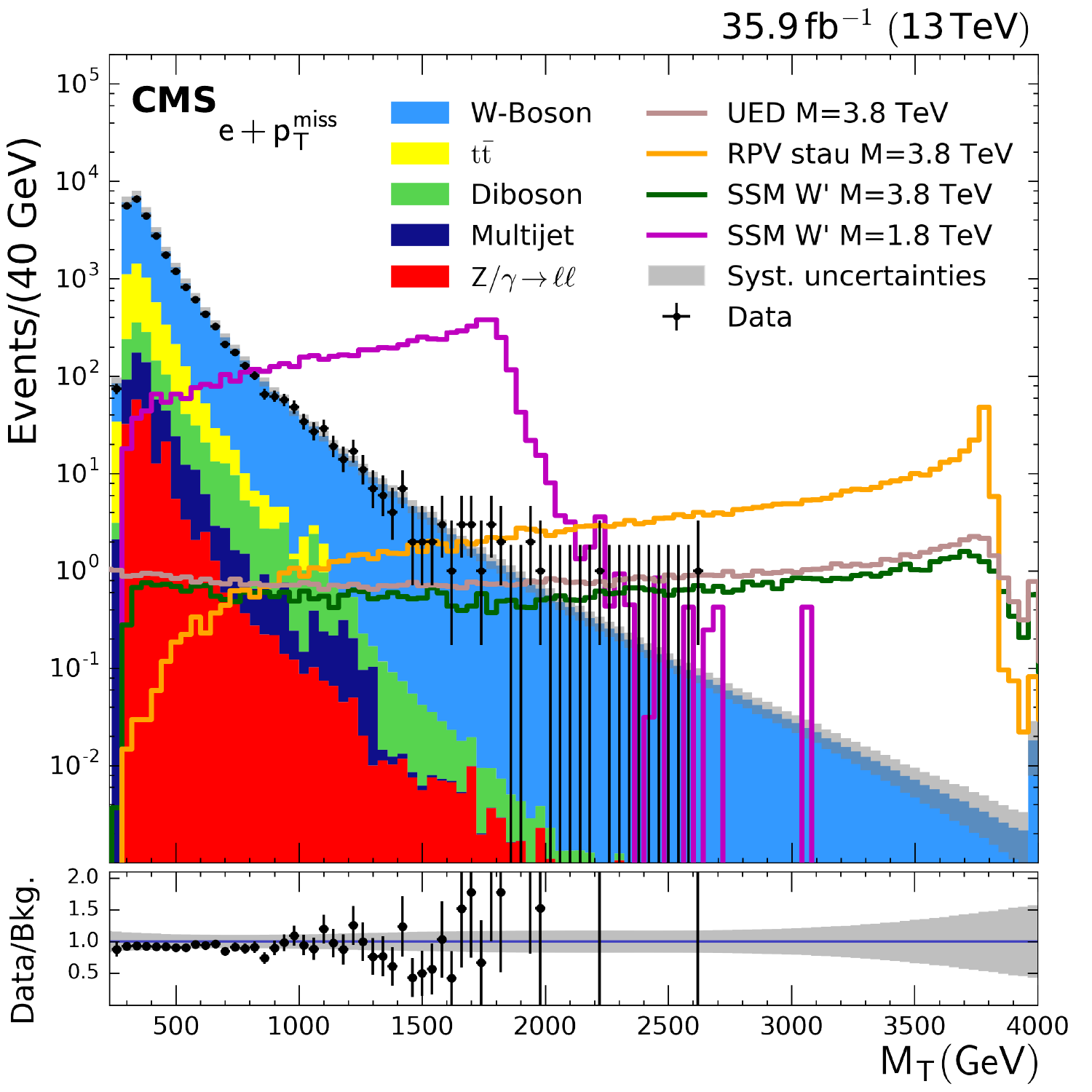}
 \includegraphics[width=0.49\textwidth]{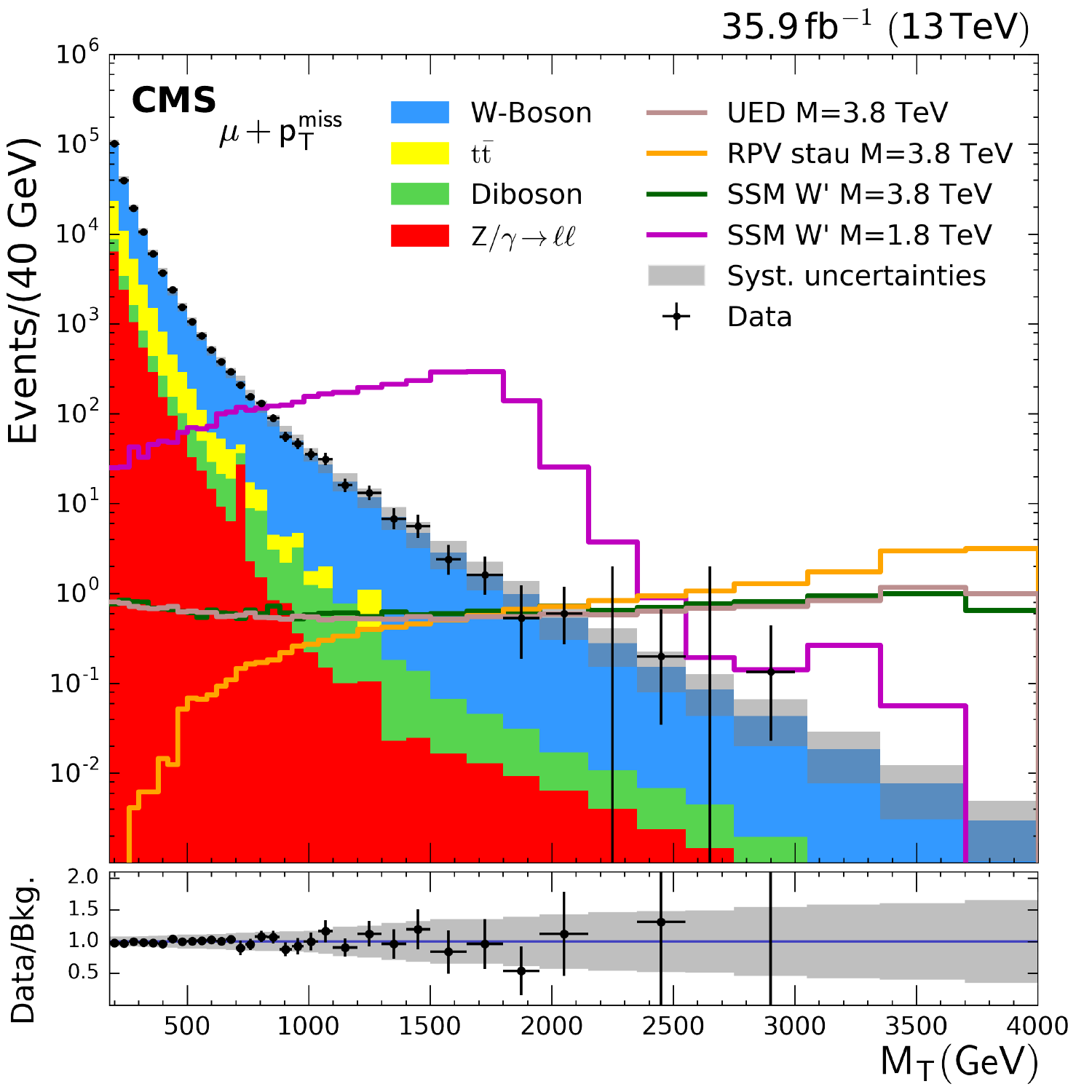}
\caption{Observed \MT distributions for the electron (left) and muon (right) channels after all selections.
Signal examples for \PWpr masses of 1.8 and 3.8\TeV, including detector simulation,
are shown in both channels. In addition, signal examples for RPV SUSY and split-UED are depicted.
The lower panel shows the ratio of data to prediction and the hatched band
includes the systematic uncertainties.
The last bin shows the total overflow.
}
\label{fig:MT}
\end{figure}

The resulting \MT distributions for the analyzed data sets are shown in Fig.~\ref{fig:MT} for
the electron and muon channels.
The minimum value of \MT is determined by the trigger thresholds, resulting in a choice of $\sim$250 and $\sim$100\GeV
in the electron and muon channels, respectively.
Included in Fig.~\ref{fig:MT} are the predicted \MT distributions
for the accepted SM events, separated into contributions from each background process, along with example signal distributions
for the SSM \PWpr, split-UED, and SUSY models.
  For the muon channel, a variable binning commensurate with the energy-dependent
\MT-resolution is used.
The expected systematic uncertainties in the predicted \MT distributions are also shown.
The numbers of signal and background events for a selected set of \MT thresholds are shown in Table~\ref{tab:Num_evts}.

No significant excess over the SM expectation is observed in the \MT spectrum.
The highest transverse mass events observed have $\MT\sim2.6$ and $\MT\sim2.9\TeV$ in the electron and muon channels, respectively.

\begin{table}[ht]
\centering
\topcaption{Expected and observed numbers of signal and background events, for a selected set of \MT thresholds.
Also shown are the total systematic uncertainties in the estimate of the event numbers.
The signal yields are based on NNLO cross sections.}
\label{tab:Num_evts}
\begin{tabular}{l c c c c} \hline
                      &   $\MT>1\TeV$ &  $\MT>2\TeV$  & $\MT>3\TeV$  & $\MT>4\TeV$ \\ \hline
Electron data         &    200        & 2             & 0            &  0 \\ [\cmsTabSkip]
Sum of SM backgrounds &    $213\pm28$  & $5.00\pm0.96$  & $0.260\pm0.077$ &  $0.0163\pm0.0078$ \\ [\cmsTabSkip]
SSM \PWpr   M = 1.8\TeV & $5040\pm770$ & $25.9\pm5.8$ & $0.43\pm0.44$   &  $0\pm0 $ \\
\hspace{1.45cm}M = 2.4\TeV & $1180\pm200$  & $560\pm100$   & $1.14\pm0.44$   &  $0\pm0 $ \\
\hspace{1.45cm}M = 3.8\TeV &  53 $\pm$ 13   & $40\pm11$    & $23.9\pm8.4$    &  $0.44\pm0.25$ \\
\hspace{1.45cm}M = 4.2\TeV &  $23.3\pm7.3$   & $17.6\pm6.5$ & $11.8\pm5.4$    &  $3.4\pm2.2$ \\
[\cmsTabSkip]
Muon data             &   208               &  4          & 0             & 0 \\ [\cmsTabSkip]
Sum of SM backgrounds &  217 $\pm$ 20     & $6.0\pm1.2$ & $0.27\pm0.21$ & $0.02\pm0.02$ \\ [\cmsTabSkip]
SSM \PWpr M = 1.8 \TeV & 5345 $\pm$ 530   & $96\pm14$  & $2.5\pm1.2$   & $0\pm0 $ \\
\hspace{1.4cm} M = 2.4 \TeV & 1282 $\pm$ 120   & $577\pm85$ & $2.4\pm1.2$   & $0.10\pm0.05 $ \\
\hspace{1.4cm} M = 3.8 \TeV &  57  $\pm$ 6     & $42\pm6$   & $24\pm12$      & $2\pm1$ \\
\hspace{1.4cm} M = 4.2 \TeV &   25 $\pm$ 3     & $19\pm3$    & $12\pm6$      & $3.6\pm1.8 $ \\
\hline
\end{tabular}
\end{table}

\section{Background}
\label{sec:background}

Searching for deviations from the steeply falling \MT spectrum requires an accurate background
estimate at very high transverse masses.
For the majority of background sources, the estimate is determined from simulation,
based on samples with large event counts at high \MT.
The primary source of background for all signals is the presence of off-peak,
high transverse mass tails of the SM $\PW \to \ell\nu$ decays. Other backgrounds arise from multijet events from QCD processes,
\ttbar and single top quark production, and from Drell-Yan events. Contributions from dibosons ($\PW\PW$, $\PW\PZ$, $\PZ\PZ$) decaying to $\Pe$ or $\mu$ are also considered.

The dominant and irreducible background is \PW $\to \ell \nu$ with $\ell= \Pe,\mu$, and $\tau$.
The $\PW\to \tau \nu$ process mostly contributes to the
region of lower \MT values in comparison to decays into the other lepton channels, because of the large fraction of the momentum carried
away by the two neutrinos from the tau lepton decay. To estimate the dominant SM \tPW boson background, several different
$\PW\to \ell \nu$ samples are used: one is generated at next-to-leading order (NLO) by \MGvATNLO~\cite{Alwall:2014hca} with
the NNPDF3.0 PDF set~\cite{Ball:2017nwa},
describing the events with an off-shell $\PW$ boson mass up to 100\GeV,
and additional samples, covering the
boson high-mass region (from 100\GeV onwards), are generated at LO with \PYTHIA~8.212,
tune CUETP8M1 and NNPDF2.3 PDF.
Higher-order electroweak (EW) and QCD multijet corrections are evaluated in bins of \MT, following the procedure used in
a previous CMS publication~\cite{EXO-15-006}.
They are calculated using {\FEWZ}~$3.1$~\cite{fewz, Li:2012wna}
at NNLO QCD multijet precision and {\sc mcsanc}~1.01~\cite{Bondarenko:2013nu}
at NLO electroweak precision.
The resulting K-factors depend on \MT, being around 1.1 for
\MT=400\GeV and decreasing to around 0.8 for \MT=3\TeV.

Top quark pair and single top quark production are other sources of high-\pt leptons and \ptmiss, and these are generated with \POWHEG
2.0~\cite{powheg_1,powheg_2,powheg_3,Re:2010bp}
in combination with \PYTHIA~8.212,
except for the $s$-channel of single top quark production, which is generated with \MGvATNLO
in combination with \PYTHIA~8.212.
The \ttbar category includes both semileptonic and dileptonic decay modes samples.
A NNLO cross section calculation from Ref.~\cite{Czakon:2013goa} is used
to rescale the NLO predictions.
These events are largely rejected by requiring compatibility with a two body decay,
but the remaining events can extend into the region of high \MT, as seen in Fig.~\ref{fig:MT}.

Drell-Yan production of dileptons ($\ell = \Pe, \mu$) constitutes a background when one lepton escapes detection.
High mass Drell-Yan samples are generated with \POWHEG
at NLO, with parton showering and hadronization described by \PYTHIA, using the
CUETP8M1 tune and NNPDF3.0 PDF set.

Contributions from dibosons are derived from inclusive samples ($\PW\PZ$, $\PZ\PZ$, including all possible final states),
generated with \PYTHIA~8.2.1.2 with the tune CUETP8M1 and the NNPDF2.3 LO PDF set,
and exclusive samples
($\PW\PW\to \ell\nu \cPq\cPq$, $\PW\PW\to 4\cPq$, $\PW\PW\to \ell\ell\nu\nu$, with $\ell=\Pe,\mu,\tau$), generated with \POWHEG. The inclusive (exclusive) normalizations are scaled to NLO (NNLO)
cross sections~\cite{Gehrmann:2014fva,Cascioli:2014yka,Campbell:2011bn}.

In the electron channel, a $\gamma$+jet event sample, generated with \MADGRAPH~5$\_$MLM~\cite{MG_MLM} at LO, is used to estimate the effects of photons misidentified as electrons.

The misidentification of jets as leptons is a possible source of background for this search.
While the contribution of QCD multijet events to the muon channel is negligible,
a small contribution to the electron channel remains after event selection.
For the latter, the shape and normalization of the QCD multijet background
is derived from data and included in the final background estimate shown in Fig.~\ref{fig:MT}.

A QCD multijet template is obtained from the events in which the electron candidate fails the isolation requirement but where all other event
requirements are met.
QCD multijet template events are scaled with normalization factors from an independent control
region, defined by the requirement $\pt/\ptmiss>1.5$, where multijet production is dominating.
In this region, the ratio
of `tight' events (electron candidates that passes all requirements of a well-isolated electron) to `loose' events (all events in the region)
is measured as a function of electron \pt.
The resulting normalization factor for QCD multijet template events is
applied to non-QCD subtracted data.
This procedure results in ratios
from about 10\% for \pt=200\GeV down to 1\% for $\pt>600\GeV$, and represents the percentage of jets that are
misreconstructed as electrons.
An uncertainty of 40\% in this estimate is obtained by comparing data to predictions from simulation, and is assigned to this small background contribution.

\section{Systematic uncertainties}
\label{sec:uncertainties}

The mismeasurement of lepton energy or momentum, arising from both detector resolution and imperfect scale calibration,
will result in a smearing of the \MT spectrum.
For each source of uncertainty, shifts of ${\pm}1\sigma$ are applied to the simulated data.
The kinematic distributions of the objects ($\Pe$, $\mu$, \ptmiss) and \MT are recalculated,
and the kinematic selection is reapplied.

The systematic uncertainty in the electron energy scale is estimated to be 0.4\,(0.8)\% in the barrel (endcaps).
For the electron energy resolution uncertainty, an additional Gaussian smearing of 1.2\,(2.4)\% for the barrel (endcap) region
is applied to the simulation~\cite{Chatrchyan:2013dga} to reflect the observed behaviour of the calorimeter.

The muon transverse momentum scale uncertainty is estimated by studying the curvature of muon tracks in
different regions of $\eta$ and $\phi$, using high-\pt muons from cosmic ray data~\cite{Chatrchyan:2012xi} and dimuon events from
high-\pt \PZ boson decays from collision data, together with corresponding simulation samples.
These studies indicate the absence of a significant
curvature bias in the central $\eta$ region ($\abs{\eta}<1.2$), within an uncertainty of
0.025/\TeV. In more forward regions, especially for $\abs{\eta}>2.1$, and in particular muon $\phi$ zones, values of the scale bias
different from zero are found. The mean value of the modification
in the simulated \pt coming from these scale biases is used as an estimate of the uncertainty, for $\pt>200\GeV$. This uncertainty is propagated to \ptmiss and \MT.

The uncertainty in the muon momentum measurement derives from the smallness of curvature of tracks for high-\pt muons,
while the energy of the electrons, measured in the crystal calorimeter, has a smaller uncertainty.
The muon \pt resolution in data at high-\pt values is well reproduced by the
simulation within an uncertainty of $\pm 1(2)\%$ in the barrel (endcap) region and no additional smearing is implemented.

The overall uncertainty in the determination of \ptmiss in each event is derived from the individual uncertainties assigned to the objects (jets, $\Pe$, $\mu$, $\tau$, $\gamma$,
and unclustered energy) obtained from the PF algorithm. The contribution of each object type is varied according to its
uncertainty.
In addition, an uncertainty of 10\% in the \pt is used for the unclustered energy.
This uncertainty is propagated to the PF \ptmiss~\cite{Chatrchyan:2011tn}. The
quadratic sum of the individual uncertainties gives the overall uncertainty in the PF \ptmiss.

In addition to the cross section uncertainties,
two further contributions play a role in the background prediction uncertainties.
For the dominant $\PW$ boson production, the uncertainties associated with the method chosen to calculate the
K-factor are considered, amounting to $\pm$5\%.
The effect of even higher order corrections
is expected to be small and therefore is not considered.
For the QCD multijet background prediction, an uncertainty of 40\%
is used as described in Section~\ref{sec:background}.

The theoretical uncertainty related to the choice of PDF set was estimated using
the PDF4LHC recommended procedure~\cite{Butterworth:2015oua}.
After reweighting the background samples with different PDF sets,
the envelopes of their error bands are taken as the uncertainty.
The values increase with \MT, ranging from an uncertainty of 1\% at \MT=300\GeV to 20\% at \MT$\approx3\TeV$.

The efficiency scale factors defined in Section~\ref{sec:objects}
are assigned uncertainties that take into account the determination method and the extrapolation to high \MT.

The simulated distribution of $\Pp\Pp$ collision vertices per bunch crossing is reweighted to the
distribution measured in data.
The uncertainty due to this reweighting method is treated as the
systematic uncertainty of the pileup simulation.
The effect on the background event yield from this uncertainty is smaller than 5\%.

The estimated uncertainty in the integrated luminosity measurement is 2.5\%~\cite{LUM-17-001}.

\section{Interpretation of the results}
\label{sec:results}

No significant deviation from the SM expectation is observed in either channel and
exclusion limits on the production cross sections of the theoretical models from Section~\ref{sec:models}
are calculated.
All limits presented here are at 95\% \CL.

The signal search is performed using the binned \MT distribution, obtained after the complete event selection,
taking into account the shape and normalization of the signal and backgrounds.
As shown in Fig.~\ref{fig:MT}, signal events
are expected to be particularly prominent at the upper end of their \MT
distribution, where the expected SM background is low.
The final \MT distributions are presented in Fig.~\ref{fig:MT} for electron and muon channels,
and together with the detailed systematic
uncertainties described in Section~\ref{sec:uncertainties}.

\subsection{Statistical analysis}
\label{sec:limits}

Upper limits on the product of the production cross section and branching fraction
$\sigma_{\PWpr} \,\mathcal{B}(\PWpr \to \ell \cPgn)$,
with $\ell = \Pe$ or $\mu$, are determined using a Bayesian method~\cite{pdg} with an uniform positive prior probability
distribution for the signal cross section.
Systematic uncertainties in the expected signal and background yields are included either via nuisance parameters with log normal
prior distributions or with the shape of the distribution included through the use of a binned likelihood (multi-bin counting).
For the SSM \PWpr and split-UED models, the limits
are obtained from the entire \MT spectrum for $\MT>220\GeV$, as displayed in Fig.~\ref{fig:MT},
using the multi-bin counting method.
This procedure is performed for different values of
parameters of each signal, to obtain limits in terms on these parameters, such as the \PWpr boson mass.

To determine a model-independent upper limit on the product of the cross section
and branching fraction,
all events above a threshold \MTlower are summed.
From the number of background events, signal events, and observed data events, the cross section limit can be calculated.
When the background is low,
this method has good sensitivity, comparable to that of the multi-bin approach.
The resulting limit can be reinterpreted for other models with a lepton and \ptmiss in the final state.
One example application is given in this paper, where limits on specific RPV SUSY processes are derived.

\subsection{Model-independent cross section limit}
\label{sec:results-mi}

A model-independent cross section limit is determined using a single bin ranging
from a lower threshold on $\MT$ to infinity.
No assumptions on the shape of the signal \MT distribution have to be made
other than that of a flat product $A\epsilon$ of acceptance and efficiency as a function of \PWpr mass.
In order to determine any limit for a specific model from the model-independent limit shown here, only the model-dependent
part of the efficiency needs to be applied.
The experimental efficiencies for the signal are already taken into account, including the effect of the kinematic selection (the cuts on \pt/\ptmiss and $\Delta \phi$), the geometrical acceptance (cut on $\eta$), and the trigger threshold.

\begin{figure}[hbtp]
\centering
\includegraphics[width=0.49\textwidth]{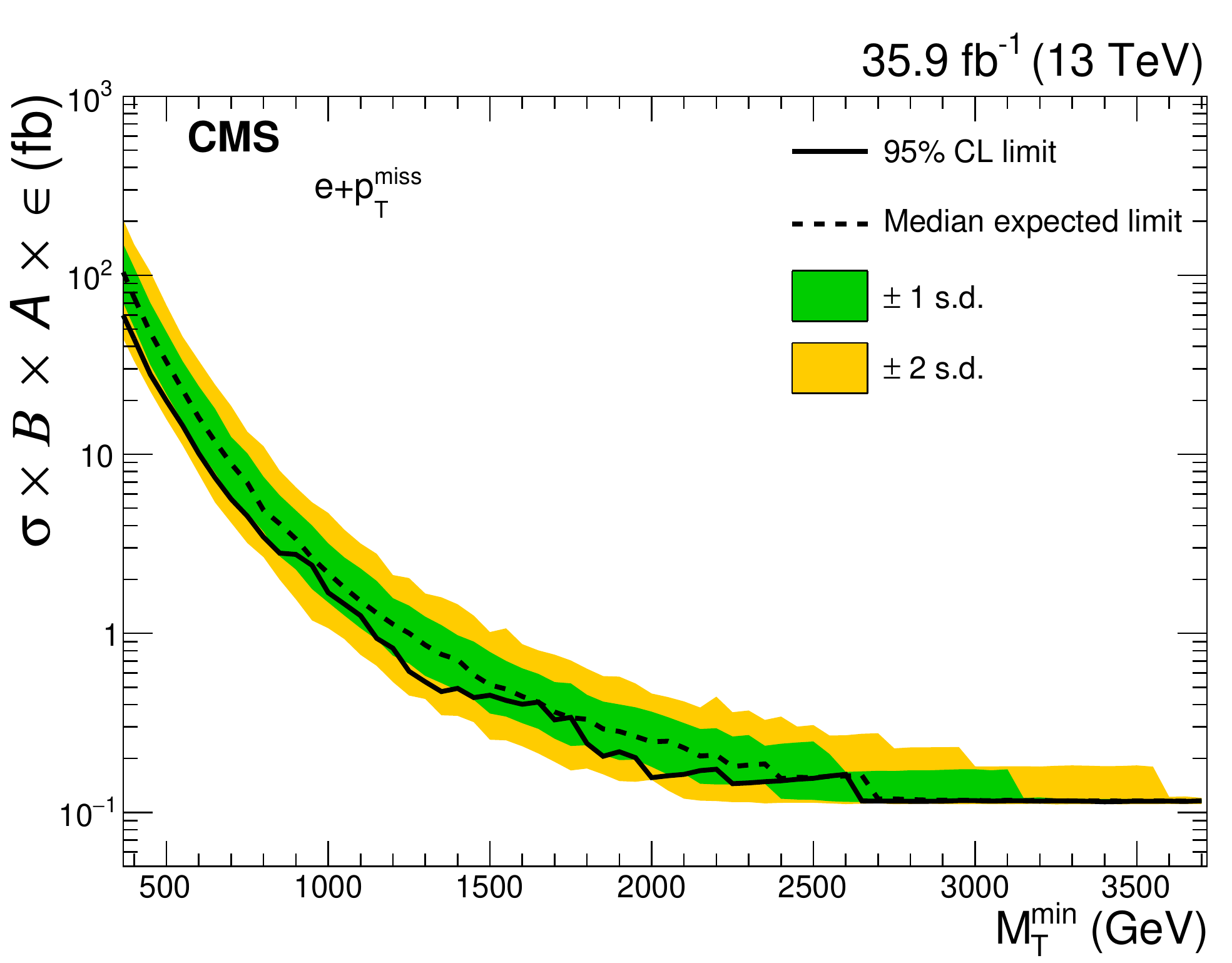}
\includegraphics[width=0.49\textwidth]{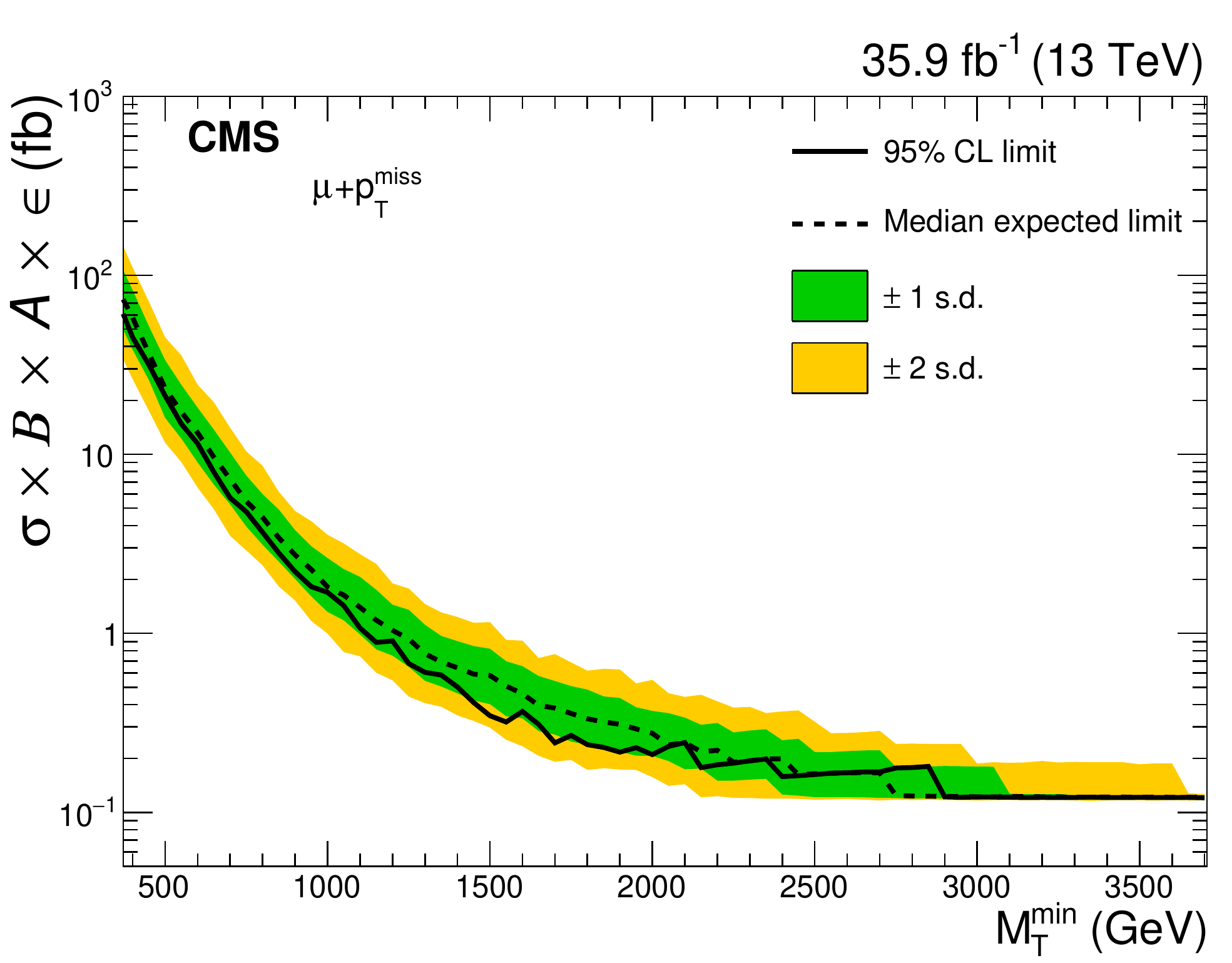}
\caption{Cross section upper limits at 95\% \CL on the effective cross section
$\sigma(\PWpr) \, \mathcal{B}(\PWpr \to \ell \nu) \, A \, \epsilon$
above a threshold $\MTlower$ for the individual electron (left) and muon (right)
channels. Shown are the observed limit (solid line), expected limit (dashed line),
and the expected limit $\pm$1 and $\pm$2 standard deviation (s.d.) intervals.
}
  \label{fig:limits-mi}
\end{figure}

A factor $f_{\MT}$ that reflects the effect of
the threshold \MTlower on the signal is determined by
counting the events with
\MT$>$\MTlower and dividing it by the number of generated events.
For $\MT>400\GeV$ the reconstruction efficiency is nearly constant,
therefore $f_{\MT}$ can be evaluated at generator level.
For lower \MT, a very slight ($<$1\%) difference is expected due to the single-lepton trigger threshold
(in particular the 130\GeV threshold for electrons).
A limit on the product of the cross section and branching fraction
$(\sigma \, \mathcal{B} \, A \, \epsilon)_\text{excl}$ can be obtained by dividing the excluded cross section
of the model-independent limit
$(\sigma \, \mathcal{B} \,  A \,  \epsilon)_\mathrm{MI}$ given in Fig.~\ref{fig:limits-mi}
by the calculated fraction $f_{\MT} (\MTlower)$:
\begin{equation*}
(\sigma \, \mathcal{B} \, A \,  \epsilon)_\text{excl} = \frac{(\sigma \, \mathcal{B} \, A \, \epsilon)_\mathrm{MI}(\MTlower)}{f_{\MT}(\MTlower)}.
\end{equation*}
Models with a theoretical cross section $(\sigma {\mathcal B})_\text{theo}$ larger than $(\sigma {\mathcal B})_\text{excl}$ can be excluded.
The procedure described here can be applied to models that exhibit back-to-back kinematics similar to those of a generic
\PWpr,
which is a reasonable assumption for a two-body decay of a massive state.
If the kinematic properties are different,
the fraction of events $f_{\MT}(\MTlower)$
needs to be determined for the model considered.
The results for the electron and muon channels are shown separately in Fig.~\ref{fig:limits-mi}.
The results depend strongly on the threshold \MTlower.
In both channels, cross section values
$\sigma \, B \, A \, \epsilon$ between 40--50\unit{fb} (for $\MTlower>400\GeV$) and 0.1\unit{fb} ($\MTlower >3\TeV$) are excluded for the \MTlower thresholds given in brackets.

\subsection{Limits on an SSM \texorpdfstring{\PWpr}{W'} boson}
\label{sec:SSMlimit}

\begin{figure}[hbtp]
\centering
 \includegraphics[width=0.49\textwidth]{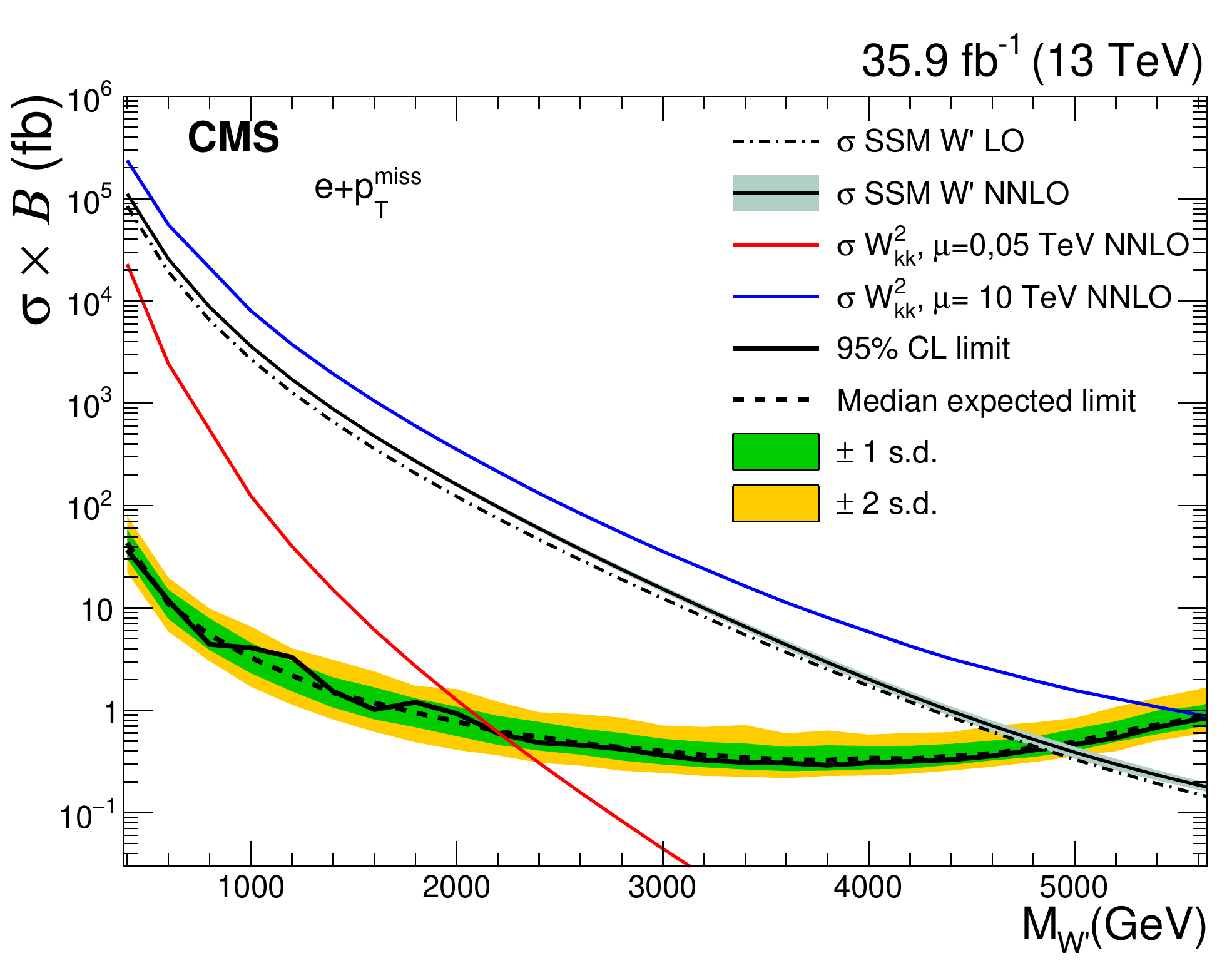}
 \includegraphics[width=0.49\textwidth]{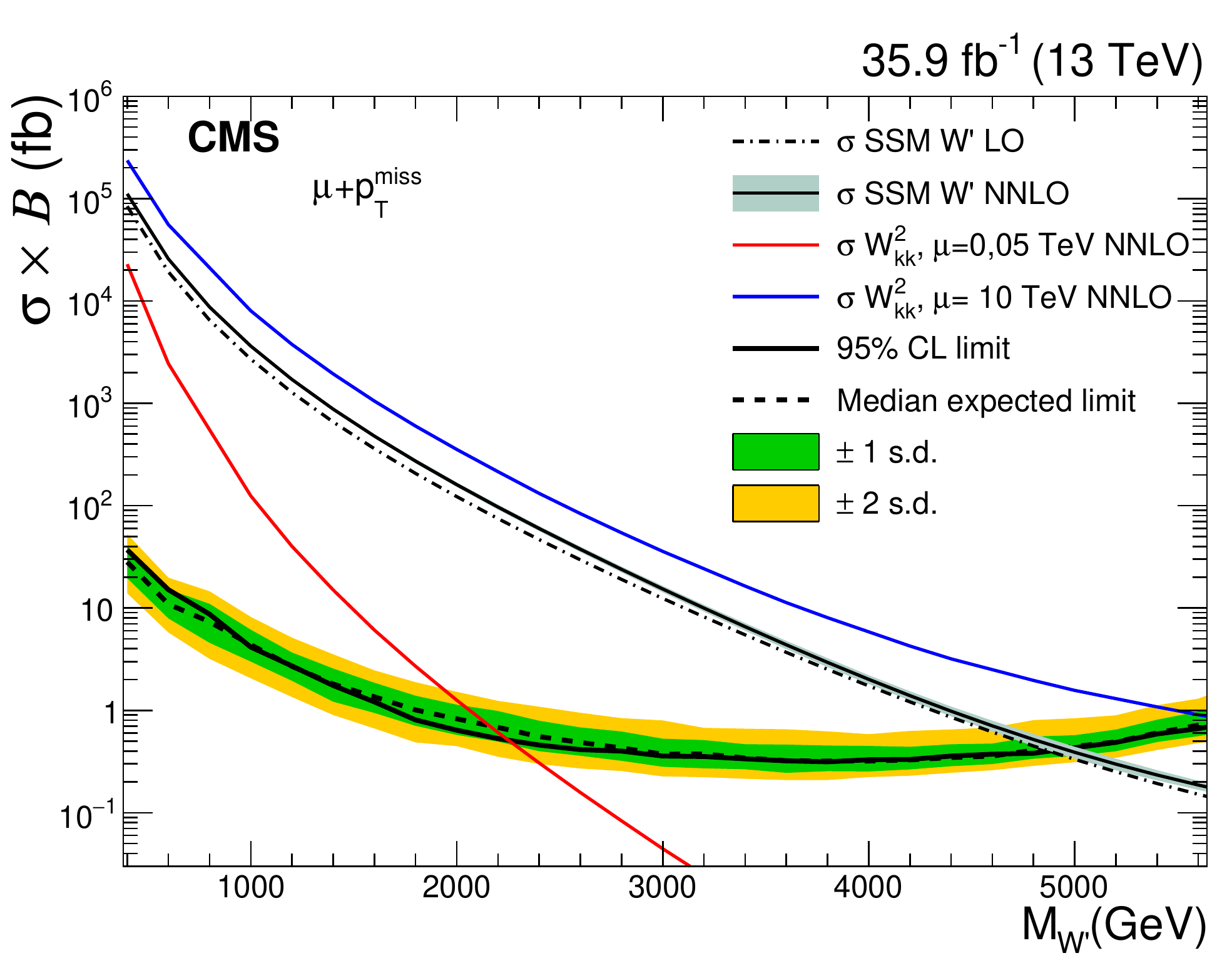}
 \caption{Expected (dashed line) and observed (solid line) 95\% \CL limits
in the SSM interpretation
for the electron (left) and muon (right) channels.
The shaded bands represent the one and two standard deviation (s.d.)
uncertainty bands.
Also shown are theoretical cross sections for the SSM benchmark model (black with a grey band for the PDF uncertainties)
and split-UED (red and blue solid lines) interpretations.
}
  \label{fig:Limit2016}
\end{figure}

The search for an SSM \PWpr boson yields limits on the product of the cross section
and branching fraction for the electron and muon channels.
The multi-bin method is used to determine the 95\% \CL upper cross section limits, shown in Fig.~\ref{fig:Limit2016}.
The indicated theoretical cross sections are the NNLO values for the lepton+\ptmiss channel, as detailed in Section~\ref{sec:models},
and are the same for both channels. The PDF uncertainties are shown as a thin band around the NNLO cross section.
The central value of the product of the theoretical cross section and branching fraction is used for deriving the mass limit.
Values of the \PWpr mass below 4.9\TeV (expected limit is 5.0\TeV) and 4.9\TeV (expected limit is 4.9\TeV)
are excluded in the electron and muon channel, respectively.
Also shown in Fig.~\ref{fig:Limit2016} are theoretical cross sections for a number of interpretations.
For the benchmark SSM model,
the displayed NNLO theoretical cross section includes the branching fraction to an electron or muon, as appropriate.

\begin{figure}[hbtp]
\centering
 \includegraphics[width=0.49\textwidth]{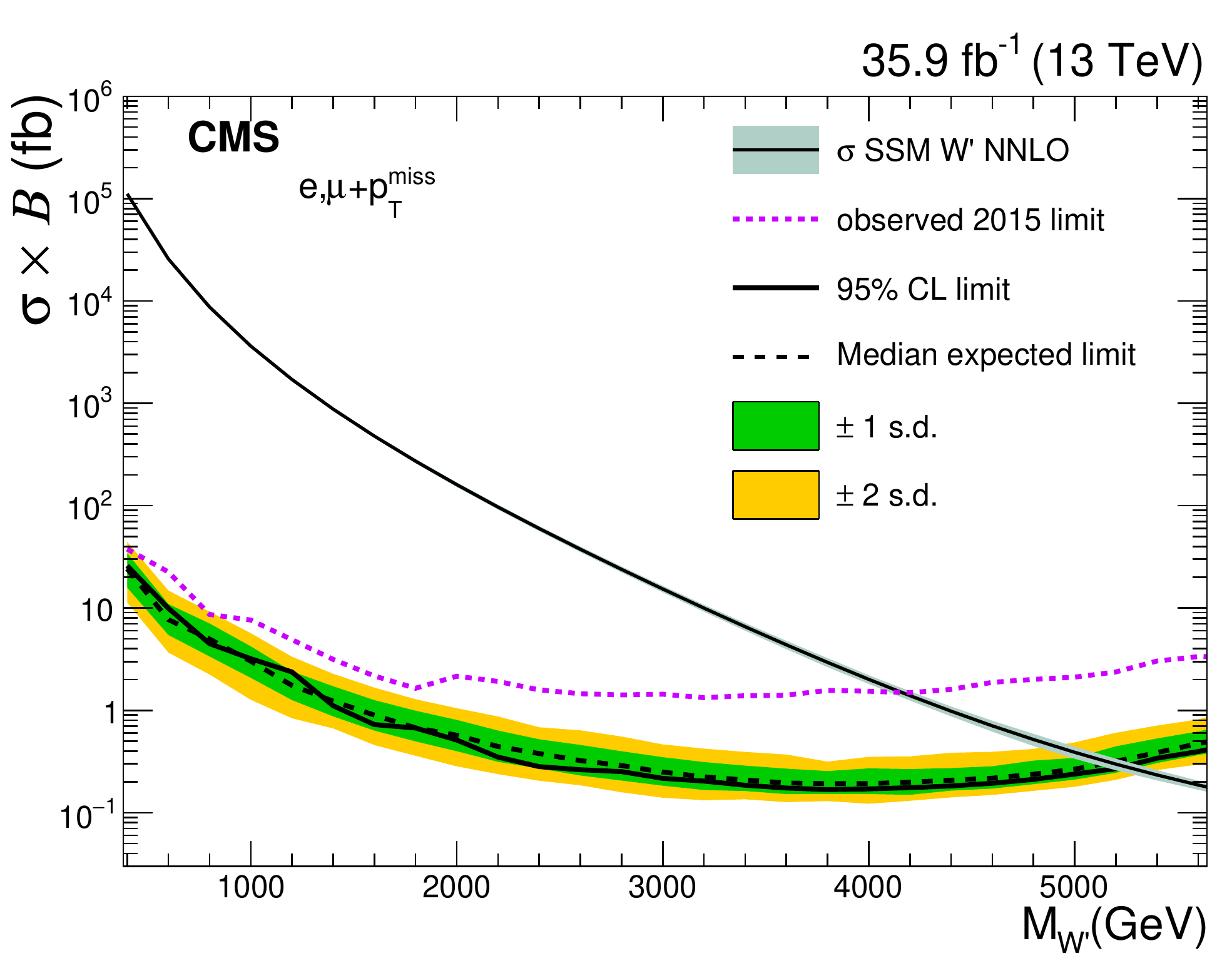}
 \includegraphics[width=0.49\textwidth]{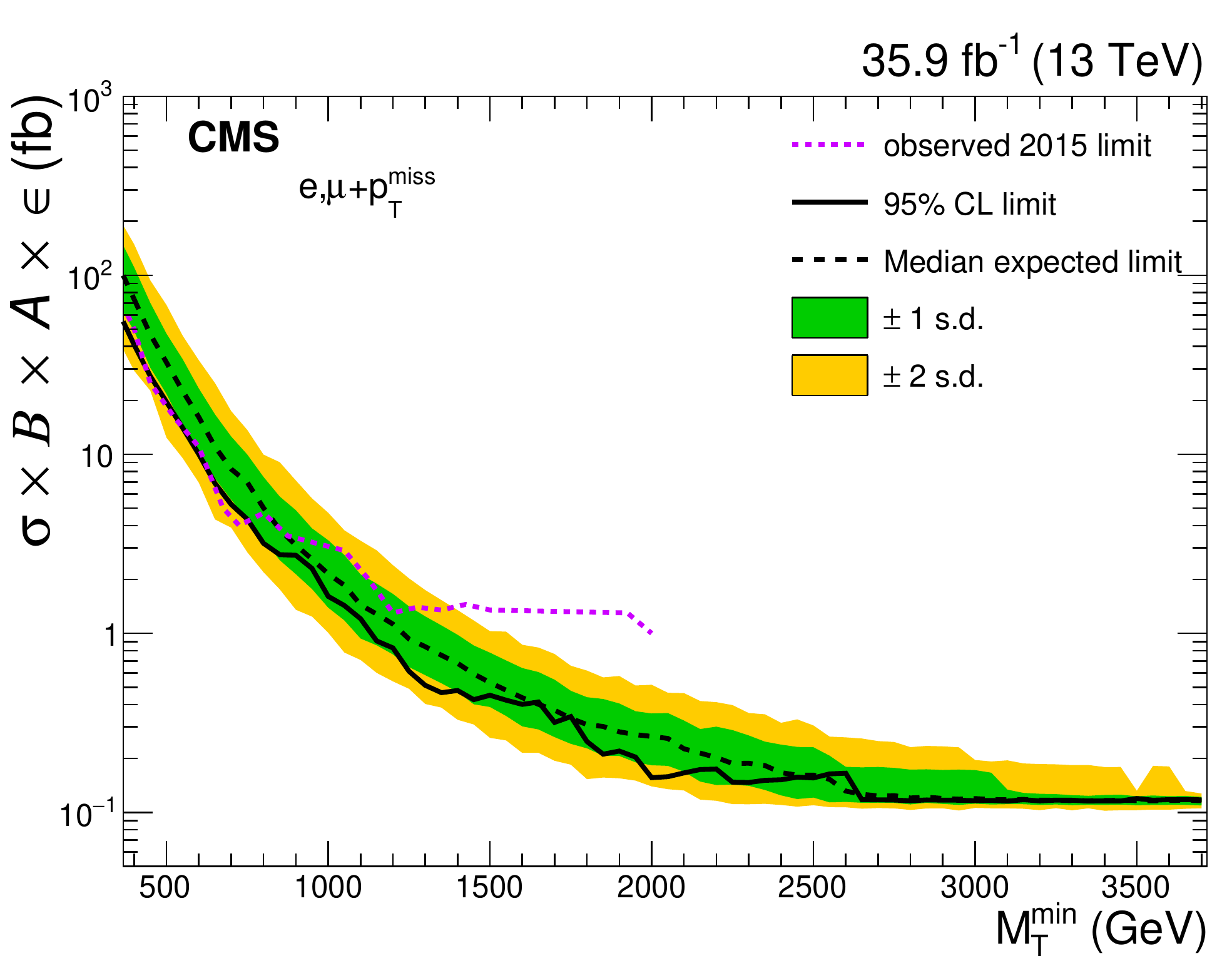}
 \caption{
Expected (dashed line) and observed (solid line)
95\% \CL limits on the product of the cross section and branching fraction, in the SSM \PWpr model (left) and the model-independent interpretation (right). Shown are the combination of the electron 
and muon channels, with the shaded bands corresponding to one and two standard deviations (s.d.). For comparison, the results from Ref.~\cite{EXO-15-006} for the regions investigated in 2015 are 
shown as dotted lines.   
}
  \label{fig:Limit2016-combined}
\end{figure}

When combining both channels, the exclusion limit
increases to 5.2\TeV (expected limit is 5.2\TeV), as depicted in Fig.~\ref{fig:Limit2016-combined}.
This is a significant improvement over the previous 13\TeV result~\cite{EXO-15-006} of 3.6\TeV and 3.9\TeV in the electron and muon channels,
respectively.
For high \PWpr masses (around 5\TeV and higher), the 	mass limit becomes less stringent because of the increasing fraction of off-shell production.
For high masses, the search sensitivity is limited by the amount of data available at present and  will improve in future.
The one and two standard deviation bands in the figures represent the systematic uncertainties as described
in Section~\ref{sec:uncertainties}.

In addition, the LO cross section for the SSM model is shown.
For the split-UED model, NNLO cross sections are displayed for two extreme values of the bulk mass parameter $\mu$
corresponding to a $\mu = 0.05\TeV$ (boson mass limit of 2.4\TeV) and a
$\mu = 10\TeV$ (boson mass limit of 5.6\TeV).

\subsection{Limits on the coupling strength}

The limit on the cross section depends on both the mass range of a potential excess
and the width. Because of the relation between the coupling
of a particle and its width, a limit can also be set on the coupling strength.

In order to compute the limit for couplings other than $g_{\PWpr}/g_{\PW}\neq1$,
reweighted samples are used that take into account the appropriate signal width
and the differences in reconstructed \MT shapes.
For $g_{\PWpr}/g_{\PW} = 1$ the theoretical LO cross sections apply.

\begin{figure}[hbtp]
\centering
 \includegraphics[width=0.49\textwidth]{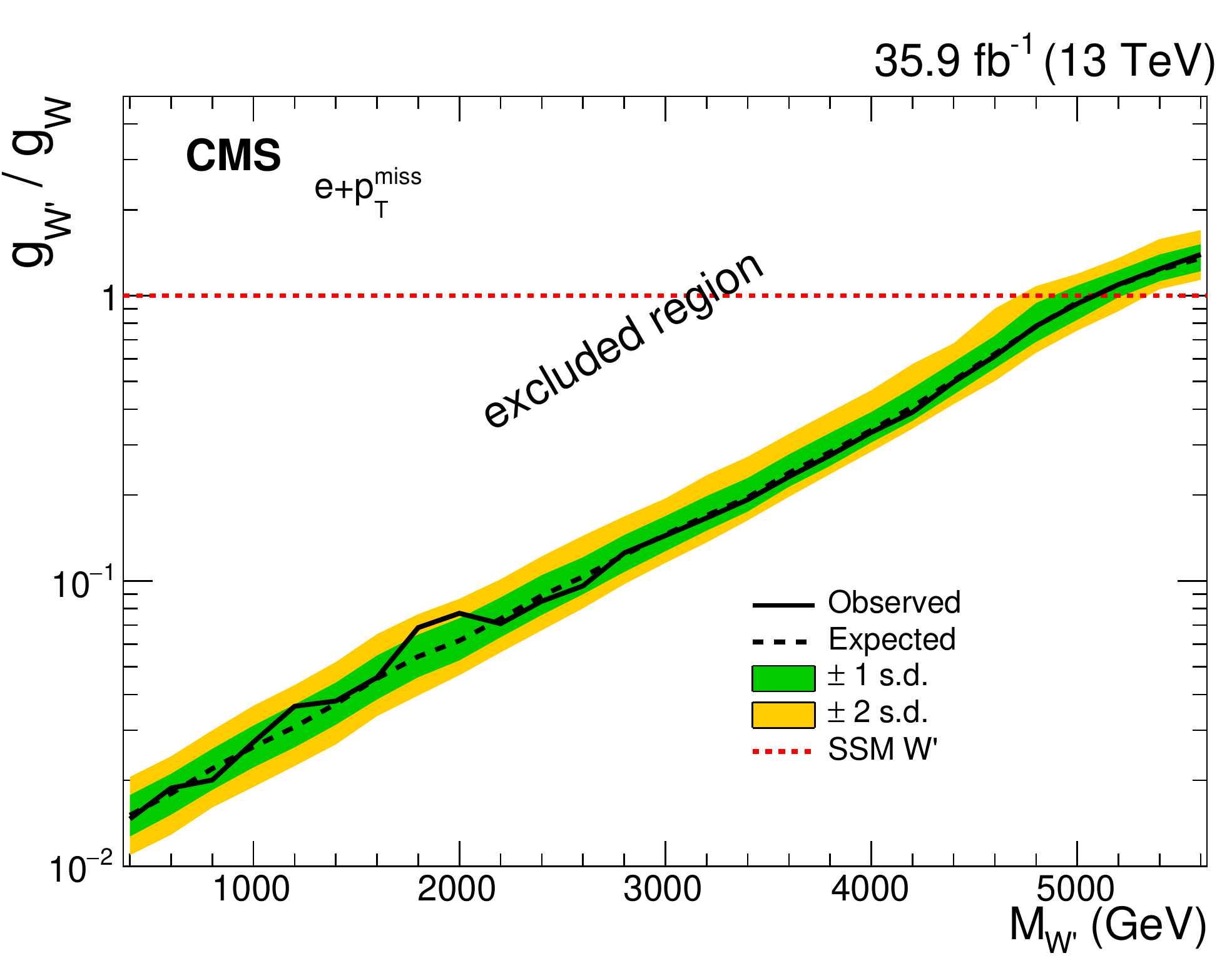}
 \includegraphics[width=0.49\textwidth]{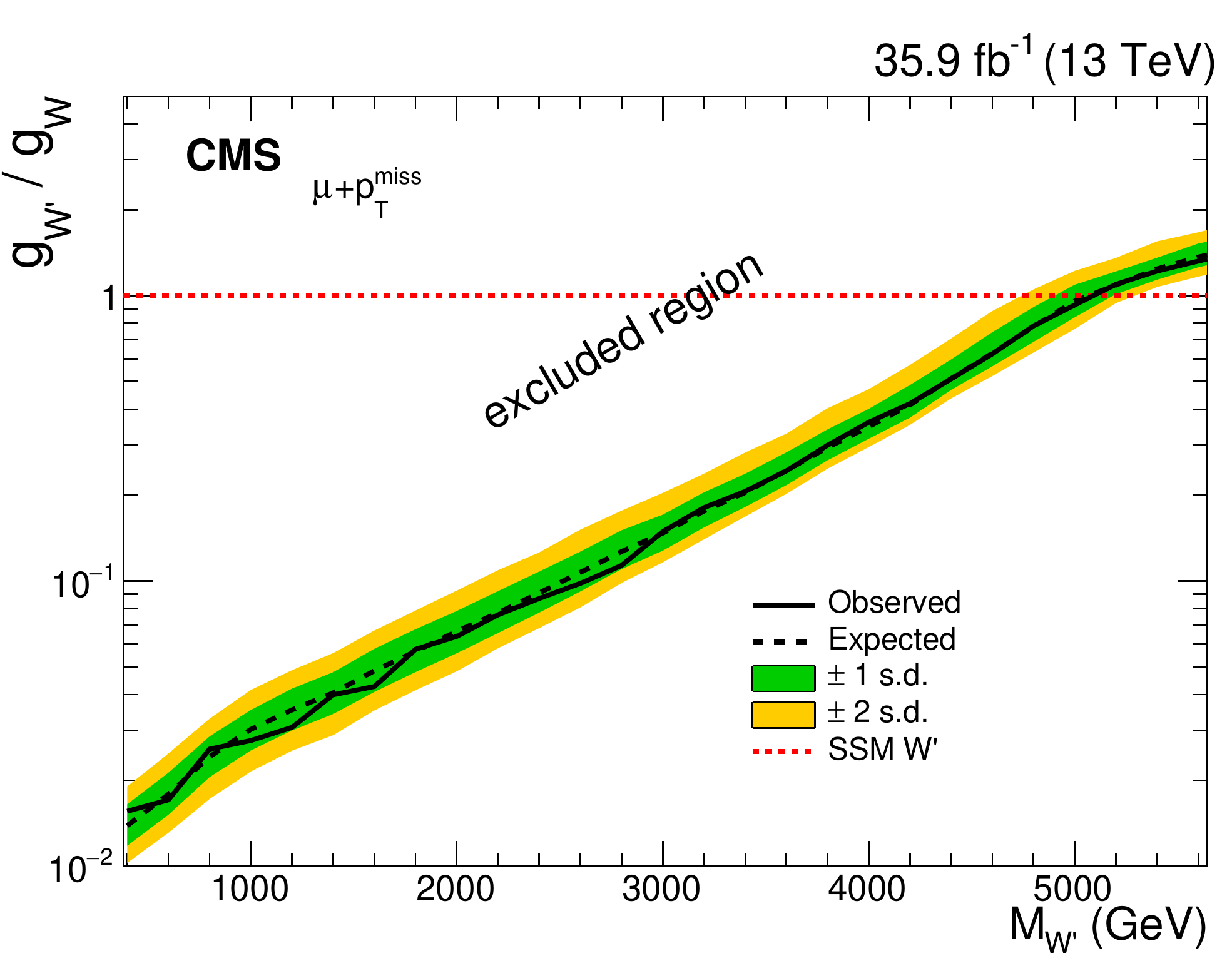}
 \caption{Expected (dashed line) and observed (solid line) 95\% \CL limits on the coupling strength
$g_{\PWpr}/g_{\PW}$ as a function of the \PWpr mass,
for the electron (left) and muon (right) channels.
The area above the limit line is excluded.
The shaded bands represent the one  and two standard deviation (s.d.) uncertainty bands.
The SSM \PWpr couplings are shown as a dotted line.
}
  \label{fig:limits-gc2016}
\end{figure}

Based on the \MT distribution of data after all selections, shown in Fig.~\ref{fig:MT},
a cross section limit as a function of the coupling is determined for each mass point.
The procedure is repeated for the full range of \PWpr masses and the corresponding intersection points provide the input for the exclusion limit on the coupling strength $g_{\PWpr}/g_{\PW}$ as a function of the \PWpr mass, as shown in Fig.~\ref{fig:limits-gc2016}.
Everything above the experimental limit is excluded.
For low masses, weak couplings down to nearly $10^{-2}$ are excluded.

\subsection{Interpretation in the split-UED model}

\begin{figure}[hbtp]
\centering
 \includegraphics[width=0.3\textwidth]{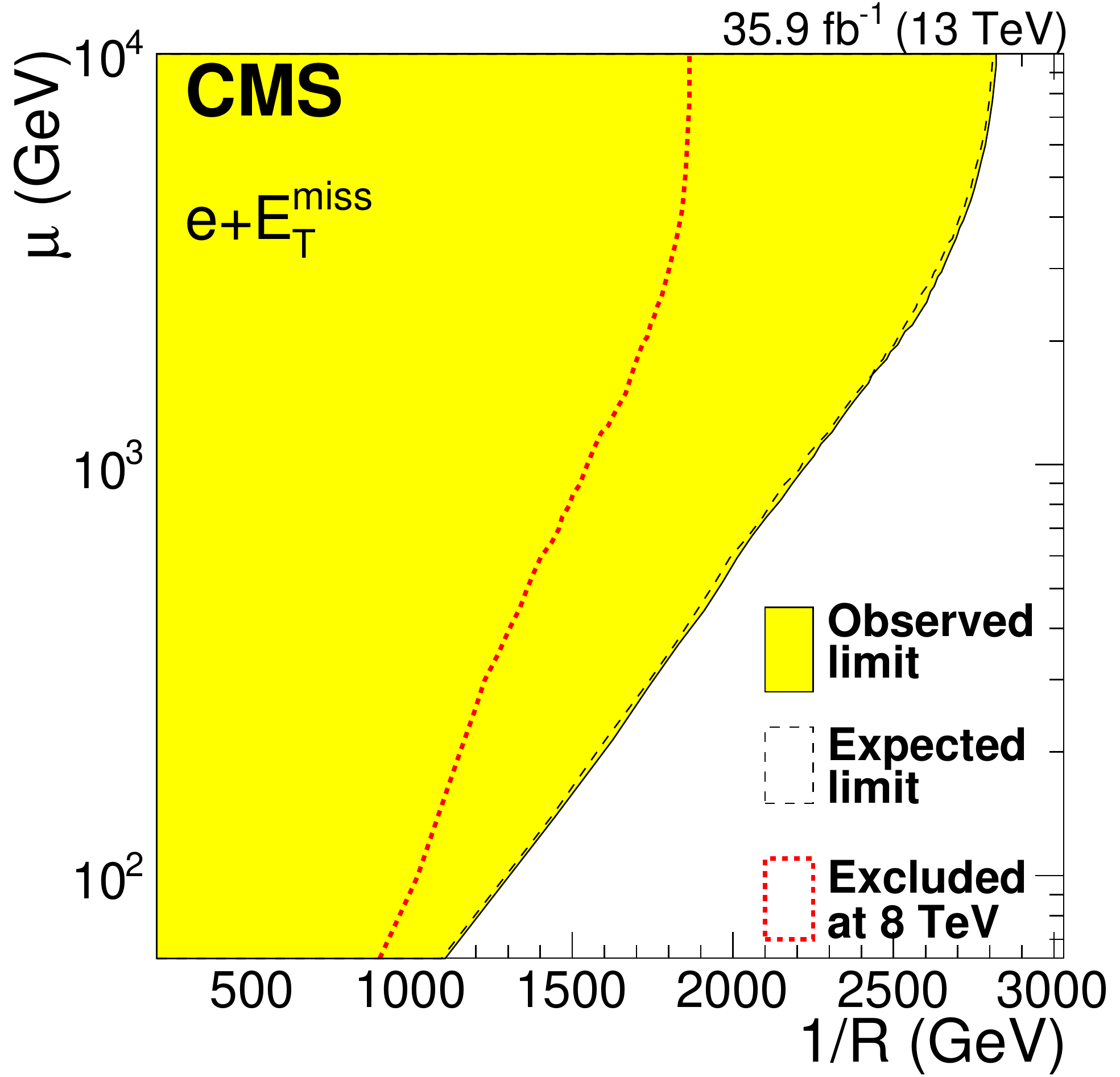}
 \includegraphics[width=0.3\textwidth]{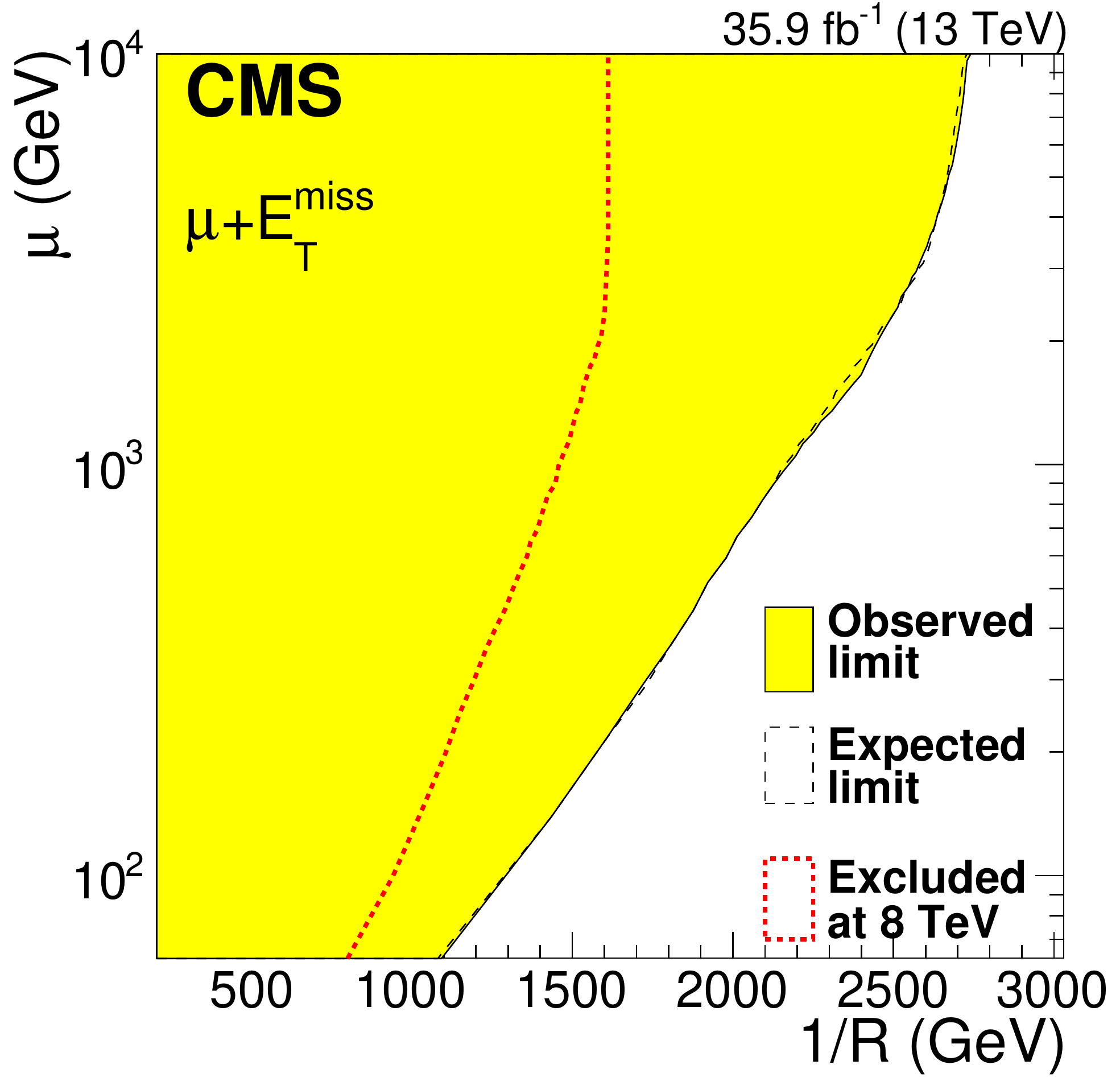}
 \includegraphics[width=0.3\textwidth]{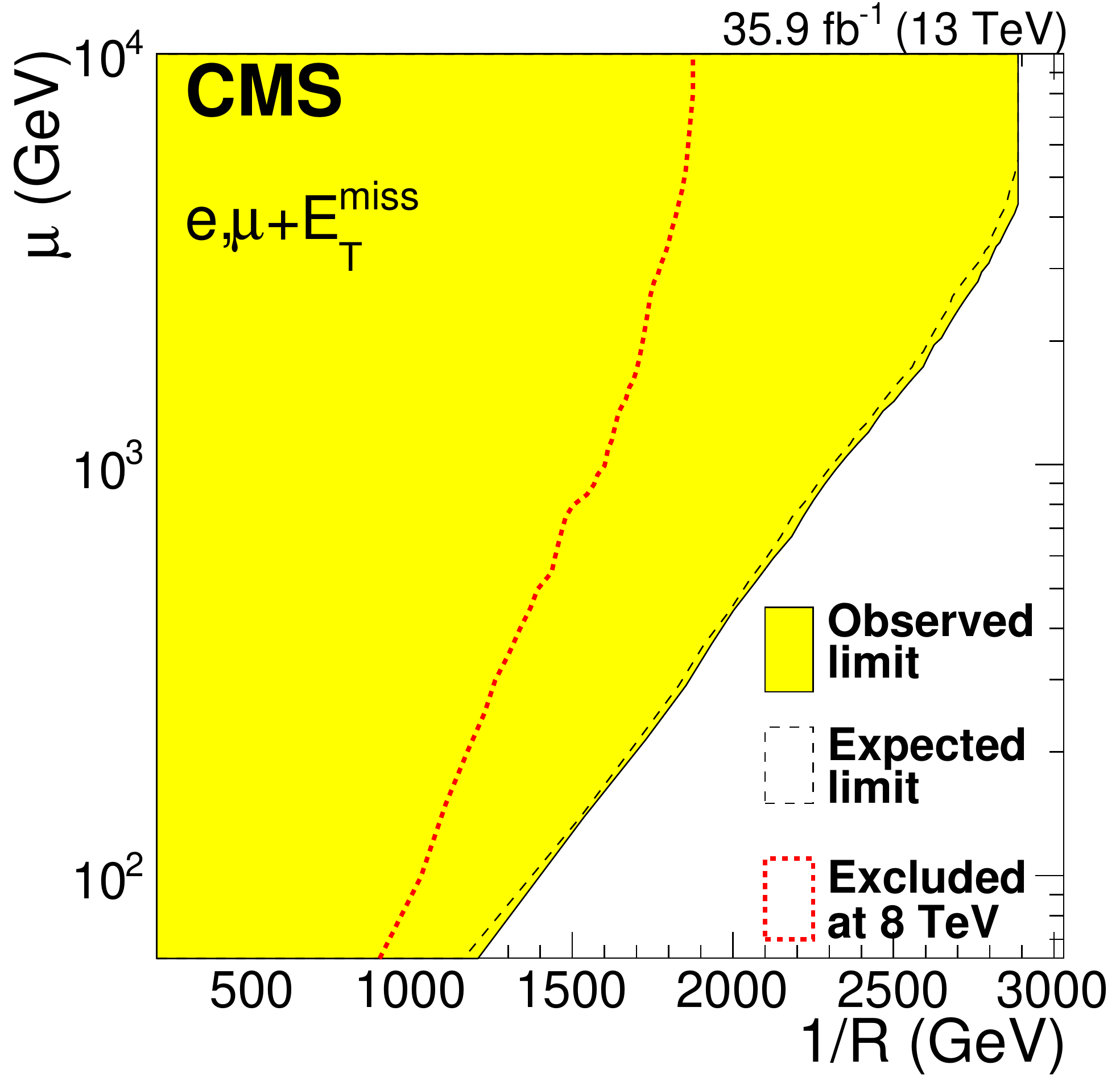}
  \caption{Exclusion limit in the plane ($\mu$, 1/R) for the split-UED interpretation for the $n=2$ case in the electron (left) and muon (middle) channels, respectively.
On the right, the result for the combination of both channels is shown.
The expected limit is depicted as a dashed line and the experimentally excluded region
as a solid black line (filled as a yellow area).
For comparison, the 8\TeV result from Ref.~\cite{EXO-12-060} is given as a red, dotted line.
}
  \label{fig:limits-ued}
\end{figure}

The UED model is parameterized by the quantities $R$ and $\mu$, which are the radius of the extra dimension and the bulk mass
parameter of the fermion field in five dimensions.
The lower limits on the mass for $n=2$ can be directly
translated from the SSM \PWpr limit into bounds on the split-UED parameter space ($1/R, \mu$),
as the signal shape
and signal efficiency are identical to the SSM \PWpr signal.
The split-UED limits are displayed in
Fig.~\ref{fig:limits-ued}, separately for each channel as well as the combination.
The observed experimental limits on the inverse radius of the extra dimension $R$ are 2.8\TeV (electrons) and 2.75\TeV (muons)
at $\mu \approx10\TeV$.
For the combination, a limit on $1/R$ of 2.9\TeV is obtained for $\mu \geq4\TeV$.
For comparison, the observed limit on $1/R$ from the 8\TeV analysis~\cite{EXO-12-060}
based on the combination of electron and muon channels is 1.8\TeV.

\subsection{Limits on RPV SUSY}

\begin{figure}[hbtp]
\centering
 \includegraphics[width=0.49\textwidth]{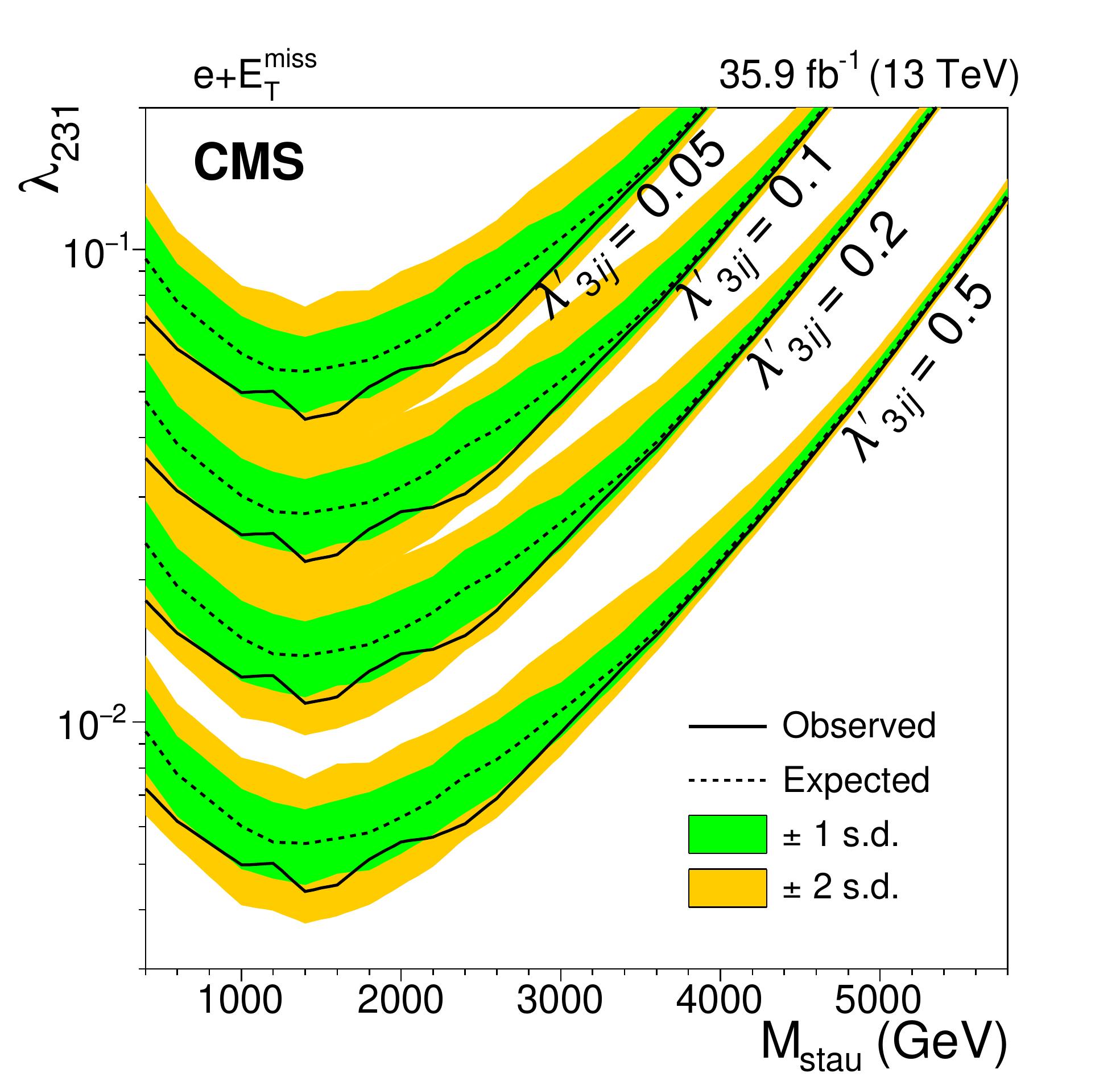}
 \includegraphics[width=0.49\textwidth]{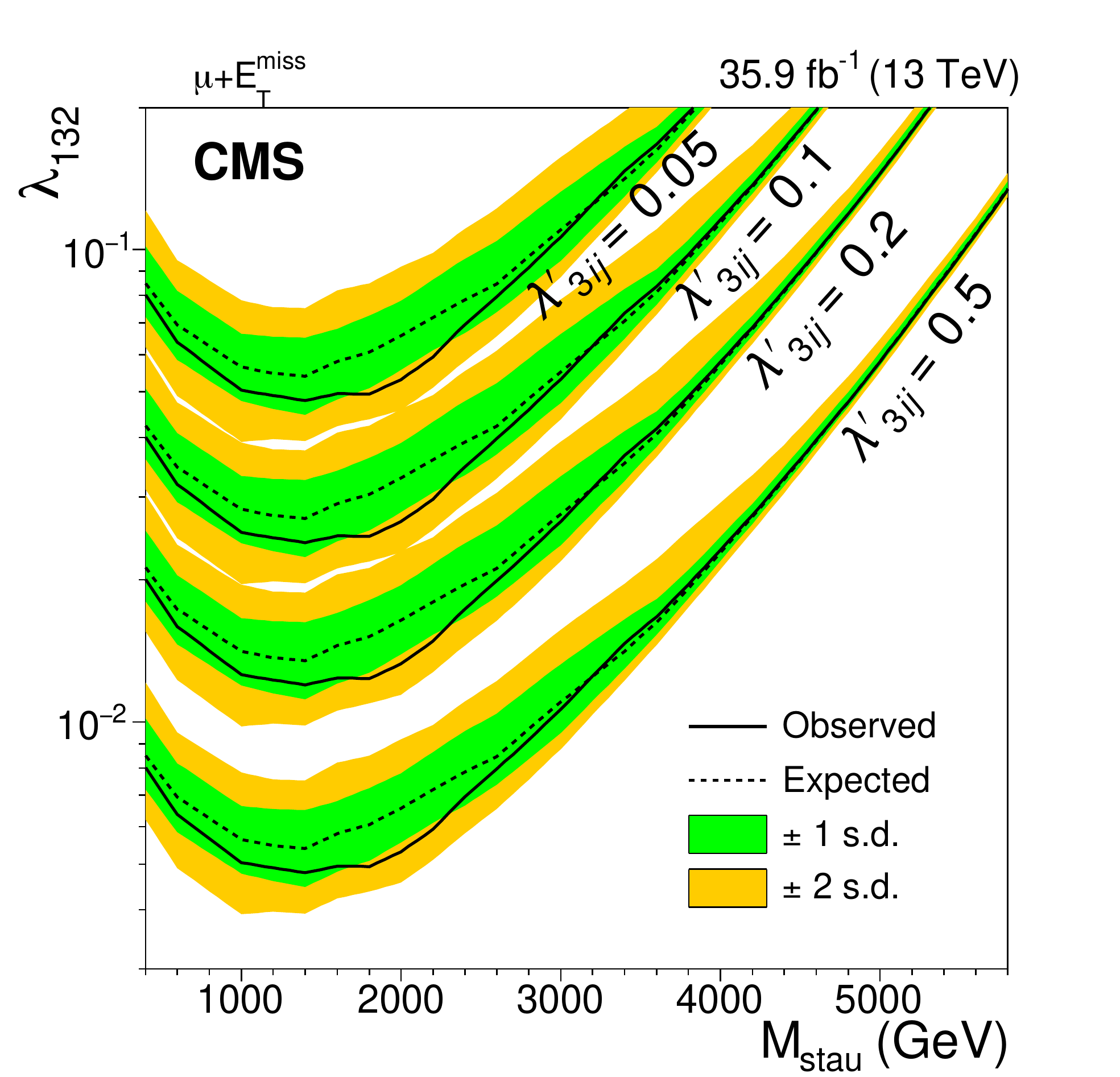}
  \caption{Observed (solid line) and expected (dashed line) exclusion limits for different couplings in the model with a $\tau$ slepton as a mediator.
The couplings $\lambda^{'}_{3ij}$, $\lambda_{231}$, and $\lambda_{132}$
are defined in Fig.~\ref{fig:feynman}.
Results are shown for the final states consisting of $\Pe+\nu_{\mu}$ on the left
and $\mu+\nu_{\Pe}$ on the right.
}
  \label{fig:limits-stau}
\end{figure}

An alternative interpretation assumes a $\tau$ slepton as a mediator, with distinct RPV SUSY
couplings at the production and decay vertices, as detailed in Section~\ref{sec:models-rpv}.
This result is obtained using the model-independent limit, and illustrates its power.
The couplings are defined in Fig.~\ref{fig:feynman}.
While $\lambda^{'}_{3ij}$ is common to both decay channels, the coupling at the
decay is either $\lambda_{231}$ (for $\Pe+\nu_{\mu}$) or
$\lambda_{132}$ (for $\mu+\nu_{\Pe}$).
This signal has a slightly different shape compared to the SSM \PWpr, a sharper Jacobian peak
and essentially no off-shell part at high mediator masses affecting the signal efficiency.
In particular at low masses, the tau slepton signal efficiency is higher.
Applying the procedure from Section~\ref{sec:results-mi}, the fraction of events $f_{\MT}$
is determined for the $\tau$ slepton model at
generator level and a correction with respect to the SSM \PWpr
is applied.
For a $\tau$ slepton mass of 3.6\TeV,
as a function of \MTlower, representative values are: $f_{\MT}$ is 0.99 ($400\GeV<\MTlower<750\GeV$), 0.90 ($\MTlower = 1650\GeV$),
and 0.82 ($\MTlower  =  2100\GeV$).
Exclusion limits on $\lambda_{231}$ and $\lambda_{123}$ as a function of the mediator mass for a number of coupling values
$\lambda^{\prime}_{3ij}$ are shown in Fig.~\ref{fig:limits-stau}.

\section{Summary}
\label{sec:summary}

A search for high-mass resonances in the lepton plus the missing transverse momentum final state
in proton-proton collisions at $\sqrt{s} = 13\TeV$ has been performed,
using a data sample corresponding to an integrated luminosity of 35.9\fbinv.
No evidence for new physics is observed when examining the transverse
mass distributions in the electron and muon channels. These observations are interpreted as
95\% confidence limits on the parameters of several models.

The exclusion limits on a sequential standard model-like \PWpr are calculated to be
4.9 (4.9)\TeV in the individual electron (muon) channels.
A combination of both channels increases the limit to 5.2\TeV, assuming standard model couplings.
Additionally,
variations in the coupling strength are studied and
couplings above $2 \times 10^{-2}$ are excluded for low \PWpr masses.
These results are also applied to the split universal extra dimension
model, and the inverse radius of the extra dimension $1/R$ is constrained by this
analysis to be above 2.9\TeV.

These results are presented in a model-independent form, making possible their interpretation in a number
of other models. An example of this application is given using a
supersymmetric model with R-parity violation, and a tau slepton mediator
with flavor-violating decays into either $\Pe+\nu_{\mu}$ or $\mu+\nu_{\Pe}$.
Limits on the coupling strengths at the decay vertex have been derived as a function of the mediator mass,
for various values of the coupling  at the production vertex
$\lambda^{'}_{3ij}$.

\clearpage

\begin{acknowledgments}

We congratulate our colleagues in the CERN accelerator departments for the excellent performance of the LHC and thank the technical and administrative staffs at CERN and at
other CMS institutes for their contributions to the success of the CMS effort. In addition, we gratefully acknowledge the computing centers and personnel of the Worldwide LHC
Computing Grid for delivering so effectively the computing infrastructure essential to our analyses. Finally, we acknowledge the enduring support for the construction and
operation of the LHC and the CMS detector provided by the following funding agencies: BMWFW and FWF (Austria); FNRS and FWO (Belgium); CNPq, CAPES, FAPERJ, and FAPESP (Brazil);
MES (Bulgaria); CERN; CAS, MoST, and NSFC (China); COLCIENCIAS (Colombia); MSES and CSF (Croatia); RPF (Cyprus); SENESCYT (Ecuador); MoER, ERC IUT, and ERDF (Estonia); Academy
of Finland, MEC, and HIP (Finland); CEA and CNRS/IN2P3 (France); BMBF, DFG, and HGF (Germany); GSRT (Greece); NKFIA (Hungary); DAE and DST (India); IPM (Iran); SFI (Ireland);
INFN (Italy); MSIP and NRF (Republic of Korea); LAS (Lithuania); MOE and UM (Malaysia); BUAP, CINVESTAV, CONACYT, LNS, SEP, and UASLP-FAI (Mexico); MBIE (New Zealand); PAEC
(Pakistan); MSHE and NSC (Poland); FCT (Portugal); JINR (Dubna); MON, RosAtom, RAS, RFBR and RAEP (Russia); MESTD (Serbia); SEIDI, CPAN, PCTI and FEDER (Spain); Swiss Funding
Agencies (Switzerland); MST (Taipei); ThEPCenter, IPST, STAR, and NSTDA (Thailand); TUBITAK and TAEK (Turkey); NASU and SFFR (Ukraine); STFC (United Kingdom); DOE and NSF (USA).

\hyphenation{Rachada-pisek} Individuals have received support from the Marie-Curie program and the European Research Council and Horizon 2020 Grant, contract No. 675440
(European Union); the Leventis Foundation; the A. P. Sloan Foundation; the Alexander von Humboldt Foundation; the Belgian Federal Science Policy Office; the Fonds pour la
Formation \`a la Recherche dans l'Industrie et dans l'Agriculture (FRIA-Belgium); the Agentschap voor Innovatie door Wetenschap en Technologie (IWT-Belgium); the F.R.S.-FNRS and
FWO (Belgium) under the ``Excellence of Science - EOS" - be.h project n. 30820817; the Ministry of Education, Youth and Sports (MEYS) of the Czech Republic; the Lend\"ulet
(``Momentum") Program and the J\'anos Bolyai Research Scholarship of the Hungarian Academy of Sciences, the New National Excellence Program \'UNKP, the NKFIA research grants
123842, 123959, 124845, 124850 and 125105 (Hungary); the Council of Science and Industrial Research, India; the HOMING PLUS program of the Foundation for Polish Science,
cofinanced from European Union, Regional Development Fund, the Mobility Plus program of the Ministry of Science and Higher Education, the National Science Center (Poland),
contracts Harmonia 2014/14/M/ST2/00428, Opus 2014/13/B/ST2/02543, 2014/15/B/ST2/03998, and 2015/19/B/ST2/02861, Sonata-bis 2012/07/E/ST2/01406; the National Priorities Research
Program by Qatar National Research Fund; the Programa Estatal de Fomento de la Investigaci{\'o}n Cient{\'i}fica y T{\'e}cnica de Excelencia Mar\'{\i}a de Maeztu, grant
MDM-2015-0509 and the Programa Severo Ochoa del Principado de Asturias; the Thalis and Aristeia programs cofinanced by EU-ESF and the Greek NSRF; the Rachadapisek Sompot Fund
for Postdoctoral Fellowship, Chulalongkorn University and the Chulalongkorn Academic into Its 2nd Century Project Advancement Project (Thailand); the Welch Foundation, contract
C-1845; and the Weston Havens Foundation (USA).

\end{acknowledgments}
\bibliography{auto_generated}

\cleardoublepage \appendix\section{The CMS Collaboration \label{app:collab}}\begin{sloppypar}\hyphenpenalty=5000\widowpenalty=500\clubpenalty=5000\input{EXO-16-033-authorlist.tex}\end{sloppypar}
\end{document}

%% file: EXO-16-033-authorlist.tex
\vskip\cmsinstskip
\textbf{Yerevan Physics Institute,  Yerevan,  Armenia}\\*[0pt]
A.M.~Sirunyan,  A.~Tumasyan
\vskip\cmsinstskip
\textbf{Institut f\"{u}r Hochenergiephysik,  Wien,  Austria}\\*[0pt]
W.~Adam,  F.~Ambrogi,  E.~Asilar,  T.~Bergauer,  J.~Brandstetter,  E.~Brondolin,  M.~Dragicevic,  J.~Er\"{o},  A.~Escalante Del Valle,  M.~Flechl,  M.~Friedl,  R.~Fr\"{u}hwirth\cmsAuthorMark{1},  V.M.~Ghete,  J.~Grossmann,  J.~Hrubec,  M.~Jeitler\cmsAuthorMark{1},  A.~K\"{o}nig,  N.~Krammer,  I.~Kr\"{a}tschmer,  D.~Liko,  T.~Madlener,  I.~Mikulec,  E.~Pree,  N.~Rad,  H.~Rohringer,  J.~Schieck\cmsAuthorMark{1},  R.~Sch\"{o}fbeck,  M.~Spanring,  D.~Spitzbart,  A.~Taurok,  W.~Waltenberger,  J.~Wittmann,  C.-E.~Wulz\cmsAuthorMark{1},  M.~Zarucki
% ...
\vskip\cmsinstskip
% ...
1:~Also at Vienna University of Technology,  Vienna,  Austria\\
% ...
